\documentclass[useAMS,usenatbib]{mn2e}
\usepackage{latexsym,amsmath,amssymb,enumerate}
\usepackage[]{graphicx,lscape}
\usepackage{subfigure}
\usepackage{multirow}

\usepackage{verbatim}
\usepackage[usenames,dvipsnames]{color}

\usepackage{soul} % used for strikeout, usage \st{strike_out_word} 

\usepackage{ulem}
\usepackage[usenames,dvipsnames]{color}
\usepackage{algorithmicx,algorithm,algpseudocode}
\usepackage{mathabx}
\usepackage{listings}
\def \jeans {\textsc{JEAnS} }

\title[A novel \textsc{JEAnS} analysis of Fornax ]{A novel \textsc{JEAnS} analysis of the Fornax
dwarf using evolutionary algorithms: mass follows light with signs of an off-centre merger.}
\author[Diakogiannis et al.]{Foivos I. Diakogiannis$^{1}$\thanks{E-mail:
foivos.diakogiannis@uwa.edu.au}, Geraint F. Lewis$^{2}$, 
Rodrigo A. Ibata$^{3}$, Magda Guglielmo$^{2}$,
\newauthor  Prajwal R. Kafle$^{1}$, 
 Mark I. Wilkinson$^{4}$  and Chris Power$^{1}$\\
$^{1}$International Center for Radio Astronomy Research, University of Western Australia, 35 Stirling Highway, Crawley, WA 6009, Australia\\
$^{2}$Sydney
Institute for Astronomy, School of Physics, A28, University of Sydney, NSW 2006, Australia\\
$^{3}$Observatoire
Astronomique, Universit\'{e} de Strasbourg, CNRS, 11, rue de l Universit\'{e}, F-67000 Strasbourg, France\\
$^{4}$Department of Physics \& Astronomy, University of Leicester, Leicester LE1 7RH, UK }

\begin{document}

%\date{ {\bf CHANGE!!!} Accepted 1988 December 15. Received 1988 December 14; in original form 1988 October 11}

%\pagerange{\pageref{firstpage}--\pageref{lastpage}} \pubyear{2013}

\maketitle

\label{firstpage}
\begin{abstract}
Dwarf galaxies, among the most dark matter dominated structures of our universe, are  excellent   test-beds for dark matter theories. 
Unfortunately,  mass modelling of these systems suffers from the  well documented mass-velocity anisotropy degeneracy. 
For the case of spherically symmetric systems, we 
describe a method for  non-parametric modelling of the radial and tangential velocity moments. The method is a numerical velocity anisotropy ``inversion'', with parametric mass models, where the radial velocity dispersion profile, $\sigma_{\mathrm{rr}}^2$ is modeled as a B-spline, and the optimization is a three step process that 
consists of: (i)  an Evolutionary modelling to determine the mass model form and the best B-spline basis to represent $\sigma_{\mathrm{rr}}^2$; (ii) an optimization of the smoothing parameters; (iii) a Markov chain Monte Carlo analysis to determine the physical parameters. 
The mass-anisotropy degeneracy is reduced into mass model inference, irrespective of kinematics. 
We test our method using synthetic data. 
Our  algorithm  constructs  the best kinematic profile and 
 discriminates between competing dark matter models.
We apply our method to the Fornax dwarf spheroidal galaxy. 
Using a King brightness profile and testing various dark matter mass models, our model inference  favours   a simple mass-follows-light system. We find that the anisotropy profile of Fornax is tangential ($\beta(r) < 0$) and  we estimate a total mass of $M_{\text{tot}} = 1.613 ^{+0.050}_{-0.075} \times 10^8 \, \text{M}_{\odot}$, and a mass-to-light ratio  of 
$\Upsilon_V = 8.93 ^{+0.32}_{-0.47} \, (\text{M}_{\odot}/\text{L}_{\odot})$.  
The algorithm we present is a robust and computationally inexpensive method for non-parametric modelling of spherical clusters independent of the mass-anisotropy degeneracy.   

\end{abstract}


\section{Introduction}

Dwarf spheroidal galaxies (hereafter dSph) are some of the most  dark matter (hereafter DM) dominated structures in our universe. As such they are also some of the best laboratories for the search of DM.  
The nature of DM still eludes us: a discrepancy between theoretical predictions of  the standard $\Lambda$CDM cosmological paradigm and observations is the so called core-cusp problem. While numerical simulations predict   cuspy profiles,  predictions made from observations of dSph galaxies tend to favour cored profiles \citep{2007ApJ...663..948G, 2009MNRAS.393L..50E}. The debate is still open \citep{2015PNAS..11212249W} therefore we need methods that can potentially discriminate between these two categories. 
With respect to statistical modelling the question of cuspy or cored halos reduces to: {\it what is the best model fit (core or cusp) to given available data sets?} Appropriate  model inference  methods need to be applied for the most reliable results.

\subsection{Jeans modelling}
A popular  modelling technique for the estimation of DM profiles of dSph galaxies,  is the use of the spherically symmetric Jeans equation 
\citep{1980MNRAS.190..873B}. This method is known to suffer from the  
mass-velocity anisotropy degeneracy    \citep[\citealt{1982MNRAS.200..361B}, hereafter MAD; see also][]{2008gady.book.....B,2013degn.book.....M}. There has been extensive effort to break this degeneracy  by many authors with significant successes (e.g.  
\cite{1989ApJ...336..639B,
1992ApJ...391..531D,
2002MNRAS.333..697L,
2010MNRAS.401.2433M,
2010HiA....15...79W,
2011IAUS..271..110W,
2013MNRAS.429.3079M,
2013MNRAS.432.3361R,
2013MNRAS.428.3648I,2014MNRAS.443..598D,2014MNRAS.443..610D,2017arXiv170104833R} 
and others; 
see chap. 5 \cite{2014RvMP...86...47C} for a comprehensive review of MAD and  mass modelling of galaxies in general).

In this contribution, we build on previous work \citep{2014MNRAS.443..598D,2014MNRAS.443..610D} and present  the {\sc JEAnS}  algorithm  ({\sc Jeans Evolutionary Anisotropy-free Solver})  that performs  mass modelling with the use of the Jeans equation independent of anisotropy, $\beta(r)$, assumptions. In fact, we nowhere use $\beta(r)$ in our modelling approach. 
The method is a velocity anisotropy ``inversion'', in the sense that we calculate numerically the kinematic moments, $\sigma_{\mathrm{rr}}^2$ and $\sigma_{\mathrm{tt}}^2$, from assumptions on the mass models we use and the line-of-sight velocity dispersion, $\sigma_{\mathrm{los}}^2$, data. However, we do not  express $\sigma_{\mathrm{rr}}^2$ and $\sigma_{\mathrm{tt}}^2$ as functions of $\sigma_{\mathrm{los}}^2$ as is done e.g. in \cite{1982MNRAS.200..361B} or \cite{1992ApJ...391..531D} (among others). 
A key ingredient of our method is to represent the unknown kinematic profile, $\sigma_{\mathrm{rr}}^2$, with a B-spline curve\footnote{B-splines are flexible smooth curves that can take any geometric shape.} \citep{2014MNRAS.443..598D}.  Then we proceed in  three distinct steps. In the first step, we use evolutionary optimization to select the best mass model  and the B-spline basis for the  representation of the radial velocity dispersion, $\sigma_{\text{rr}}^2$. 
In the second step, we optimize the smoothing parameters for the flexible  B-spline representation of $\sigma_{\text{rr}}^2$ by using information from mock stellar systems.  Finally, in the third step, we perform Markov chain Monte Carlo analysis in order to determine the physical parameters of the system under investigation with their uncertainties.  Hence, we attack the problem of Jeans mass modelling in a twofold way: we model 
in a way that is independent of velocity  anisotropy assumptions,  and we discriminate between competing mass models. 
We will illustrate our algorithm by applying it to the well sampled Fornax dwarf spheroidal galaxy.

\subsection{Fornax dSph}
Fornax is a classical dSph  situated at a distance of approximately $d \approx 140$ kpc from the Sun.  The large number of measured heliocentric velocities that exist \citep{2009AJ....137.3100W}  have allowed for robust statistical inference. However it is still debatable which DM model (core or cuspy) best fits the observations. This has a significant impact on the prediction of its mass content. Various authors have presented mass estimates of Fornax, but most without following a robust model inference approach.  
\cite{2007ApJ...667L..53W} modelled Fornax with the assumption of constant anisotropy and a NFW \citep{1996ApJ...462..563N} profile. 
\cite{2007MNRAS.378..353K} used a method of  interlopers removal, verified with numerical simulations, and modelled Fornax assuming a constant anisotropy parameter with a mass-follows-light model.    
\cite{2009MNRAS.394L.102L} assumed a simple mass-follows-light model and  used   the line-of-sight kurtosis to tackle the MAD. 
\cite{2013MNRAS.429L..89A} and \cite{2011ApJ...742...20W} used a multiple stellar population  modelling method and excluded the presence of cuspy DM profiles. 
 \cite{2012ApJ...746...89J} modelled the dSph using Schwarzschild technique and determined that the best model is a cored  DM halo, however they used $\chi^2$ for their model selection which is not an appropriate model inference method \citep{hastie01statisticallearning}.  For numerical simulations based  attempts of modelling Fornax, see \cite{2041-8205-756-1-L18,2015MNRAS.454.2401B} and references therein. 

In this contribution we model Fornax  independently of the MAD, for a variety of DM mass models. We perform model selection using the modified Akaike criterion with second order bias correction and present our findings. 
\subsection{Outline of the paper}

In section \ref{DLI_III_Math_form} we describe the mathematical formulation of the modelling approach used in the \jeans algorithm. 
In section \ref{DLI_III_AlgoOverview} we describe in full detail the three steps of the {\sc JEAnS}.
In section \ref{DLI_III_Example_1} we describe the application of our method to  a synthetic data set. In section \ref{DLI_III_Fornax_modelling}
we apply our method to  the Fornax dSph and we report our results. Finally in section 
\ref{DLI_III_Conclusions} we present our conclusions and in section \ref{DLI_III_future} we describe possible future developments of our method. 

Readers who are interested only in our results on the Fornax dSph can go directly to section \ref{DLI_III_Fornax_modelling}.

\section{Mathematical formulation}
\label{DLI_III_Math_form}

In this section, we describe in detail the mathematical formulation of the various objects we use in the \jeans modelling approach.  
We start (Sec. \ref{DLI_section_Bsplines}) by giving a short description of B-spline functions that we use for the approximation of the radial velocity dispersion, $\sigma_{\mathrm{rr}}^2$.  
Then we proceed (Sec. \ref{DLI_DynModeling_label}) in detailing the  mass models as well as the necessary equations used in the \jeans modelling approach. Finally, (Sec. \ref{DLI_III_sect_stat_anal_primary}) we describe the likelihood function we use for the statistical analysis of the problem in the final step of the \jeans.

\subsection{B-spline functions}
\label{DLI_section_Bsplines}

In previous work \citep{2014MNRAS.443..598D} we have given a detailed description of B-spline functions. Here we summarize some key concepts for the convenience of the reader. 

Loosely speaking, a B-spline function\footnote{ The {\small C++} implementation of 
B-splines for this work can be found 
in https://github.com/feevos/bsplines.}, $f(x)$, is a linear combination of some constant coefficients, $a_i$, with some polynomial functions $B_{i,k}(x)$ (B-spline basis functions) of a given degree $k-1$, i.e. $f(x) = \sum_i a_i B_{i,k}(x)$.
 These basis functions are  defined in a bounded interval, $[a,b]$ over a non decreasing sequence of real numbers  $\xi_0=a \leq  \xi_1 \leq \ \cdots \leq \xi_m=b$. The elements of this sequence, $\xi_i$, are called {\it knots} and their distribution and numbers regulates the geometric shape of the B-spline basis $B_{i,k}(x)$.

The B-spline basis  form a mathematical basis that is ``as complete as possible'' for the approximation of real functions in a given interval. Then, for some suitable choice of coefficients, $a_i$, and B-spline basis, $B_{i,k}(x)$, a function $g(x)$ can be approximated by: 
\begin{equation}\label{DLI_III_gen_func}
g(x) \approx \sum_{i=1}^n a_i B_{i,k}(x)
\end{equation}
The quality of this approximation depends crucially on the choice of the basis functions $\{ B_{i,k}(x) \}$, i.e. on the order, $k$, of the basis and the choice of knot vector, $\{\xi_i \}$.  
In the following the order of the B-spline polynomials will be fixed to $k=5$, hence we define $B_i(x) \equiv B_{i,k}(x)$. The best knot vector,  $\{\xi_i \}$, will be a free parameter and will be evaluated with the use of evolutionary algorithms.

\subsection{Dynamical Modeling}
\label{DLI_DynModeling_label}

In this section we describe our choice of mass models, the Jeans formalism we follow, as well as our B-spline approximation of the radial velocity dispersion, $\sigma_{\text{rr}}^2$. In the following we use the symbol $\theta_{\star}$ for the set of stellar model parameters, and $\theta_{\bullet}$, for the set of DM model parameters.

\subsubsection{Mass models}

In order to perform robust model selection and to include models from both cuspy and cored profiles,  we use a variety of DM mass models,  specifically: 
\citet[][to allow
a cored dark matter profile suggested by many observations of spiral galaxy rotation curves,
e.g. \citealt{2001AJ....122.2396D}]{1995ApJ...447L..25B},  NFW \citep[][that represent well halo density profiles]{1996ApJ...462..563N}, Einasto \citep[][found by  \citealt{2004MNRAS.349.1039N} to represent even better the halo profiles, see also \citealt{2005MNRAS.362...95M} and \citealt{2006AJ....132.2685M}]{1989A&A...223...89E} and a simple black hole\footnote{Strictly speaking, a BH is not a DM profile, however the mathematical formalism does not change.} (hereafter BH) potential. The functional form of the mass density of each of these is: 
\begin{equation}
\rho_{\bullet}(r) =
\begin{cases}
\dfrac{\rho_{0 \bullet }}{(1+r/r_{\mathrm{s}}) [1+(r/r_{\mathrm{s}})^2]} & \mbox{Burkert}\\
   \dfrac{r_{\mathrm{s}}^3\rho_{0\bullet}}{r (r^2+r_{\mathrm{s}}^2)^{2}  } &\mbox{NFW} \\
\rho_{\mathrm{e}} \exp \{-d_n [ (r/r_{\mathrm{e}})^{1/n} -1]\} &\mbox{Einasto}\\
\dfrac{M_{\bullet}}{4 \pi r^2} \delta(r) &\mbox {Black hole}
  \end{cases}
\end{equation}
where $\delta(r)$ is the Dirac delta function and $d_n \approx 3 n - 1/3 + 0.0079/n$ \citep{2006AJ....132.2685M}. 
For our stellar model we use the mass density of the 
\cite{1966AJ.....71...64K} profile which depends on the functional form of the potential, $\Phi(r)$. See \cite{2014MNRAS.443..598D} for the exact formalism we follow as well as  \cite{2008gady.book.....B} and references therein. The total potential of the system results from the solution of the Poisson equation:
\begin{equation}
\nabla^2 \Phi = 4 \pi G \left(\rho_{\star} + \rho_{\bullet} \right) 
\end{equation}
In this way, the DM profile affects the solution of potential $\Phi$, and therefore the functional form of the stellar mass density, $\rho_{\star}$. Hence
the tracer projected mass density, $\Sigma(R)$, is influenced by the presence of DM  and this is reflected in the brightness profile of the tracer population. 
This is a very important constraint on the available range of parameters for the stellar and DM models. This is also a very good approximation for modelling Fornax, since we expect that the stellar mass represents  $\approx 10\%$ of its total mass, i.e. the stellar component has a significant contribution to the total mass distribution.

\subsubsection{The Jeans formalism}
In the Jeans equation formalism, a stellar system is assumed to be fully described by the stellar (tracer) mass density, $\rho_{\star}$, the DM mass density, $\rho_{\bullet}$, and the second order radial, $\sigma_{\text{rr}}^2$, and tangential, $\sigma_{\text{tt}}^2$, velocity moments.  The connection between these quantities and the line-of-sight velocity dispersion is given by the Jeans equation: 
\begin{equation}
\label{DLI_III_Jeans_std}
-\frac{\text{d}\Phi}{\text{d}r} = \frac{\text{d} \sigma_{\text{rr}}^2}{\text{d}r} + \left(
\frac{1}{\rho_{\star}} \frac{\text{d} \rho_{\star}}{\text{d}r} + \frac{2}{r}
\right) \sigma_{\text{rr}}^2 - \frac{\sigma_{\text{tt}}^2}{r}
\end{equation}
and the geometric definition of the projected line-of-sight velocity dispersion  \citep[][]{1982MNRAS.200..361B} 
\begin{equation}
\label{DLI_III_sigmaLOS_def}
\sigma_{\text{los}}^2 = \frac{1}{\Sigma(R)} 
\int_{r=R}^{r_{\text{tt}}} \frac{ \rho_{\star} [2 \sigma_{\text{rr}}^2 (r^2-R^2) + R^2 \sigma_{\text{tt}}^2 ] }{r \sqrt{r^2-R^2}} \text{d}r 
\end{equation}
where $\Phi$ is the total potential resulting from both the stellar and DM mass distributions. 
 
 The connection between the velocity moments and the anisotropy profile, is given through: 
\begin{equation} \label{DLI_III_beta_definition}
  \beta(r)=1 - \frac{ \sigma_{\text{tt}}^2}{2 \sigma_{\text{rr}}^2}
\end{equation}
where $\sigma_{\text{tt}}^2= \sigma_{\theta\theta}^2 +\sigma_{\mathrm{\phi\phi}}^2= 2\sigma_{\theta\theta}^2=2\sigma_{\mathrm{\phi\phi}}^2$.    
   
\subsubsection{Approximation of the radial velocity dispersion profile}

At the heart of our method lies the approximation of the radial velocity dispersion, $\sigma_{\text{rr}}^2$, with a B-spline function, i.e. 
\begin{equation} \label{DLI_III_sigma_r_representation}
\sigma_{\text{rr}}^2(r) = \sum_{i=1}^{N_{\text{coeffs}}} a_i B_i(r)
\end{equation}
The choice of the particular basis, $B_i(r)$, as well as the constant coefficients $a_i$ regulate the geometric shape of the $\sigma_{\text{rr}}^2(r)$ profile. 

Once the basis functions, $B_i(x)$, are defined,  the coefficients $a_i$ that regulate  the geometric shape of $\sigma_{\text{rr}}^2$ also regulate, in a linear way, the geometric shape of the line-of-sight velocity dispersion \citep{2014MNRAS.443..598D,2014MNRAS.443..610D}: 
\begin{equation}
\label{DLI_III_sigmaLOS}
\sigma_{\text{los}}^2 = \sum_i a_i I_{i}(R) + C(R)
\end{equation}
where
 \begin{align} \label{DLI_III_BSplines_I}
I_{i} (R) &\equiv \frac{1}{\Sigma(R)}
\int_{R}^{r_{\mathrm{t}}} \mathrm{d} r  \frac{   2 r \rho_{\star} B_i(r)    + R^2 \mathrm{d} ( \rho_{\star} B_i(r) ) / \mathrm{d} r }{\sqrt{r^2-R^2}}\\
\label{DLI_III_BSplines_C}
C(R) &\equiv\frac{1}{\Sigma(R)}\int_{R}^{r_{\mathrm{t}}} \frac{\rho(r) R^2}{\sqrt{r^2-R^2}} \frac{\mathrm{d}  \Phi(r)}{\mathrm{d} r} \mathrm{d} r
\end{align}
Eq.  \eqref{DLI_III_sigmaLOS} results from solving Eq. \eqref{DLI_III_Jeans_std} with respect to $\sigma_{\text{tt}}^2$ and substituting under the integral sign of Eq. \eqref{DLI_III_sigmaLOS_def}.
The functions $I_i(R)$ and $C(R)$ depend on the stellar, $\rho_{\star}$, and dark matter, $\rho_{\bullet}$, mass densities. In addition, $I_i(R)$ depends also on the particular B-spline basis $B_i(r)$, which is a known function. 
Since Eq. \eqref{DLI_III_sigmaLOS} is linear with respect to the unknowns $a_i$, for a given set of parameters for the stellar and DM models, there exists a unique solution, $\hat{a}_i$ for the kinematic profile. The quality of the fit, however, depends on the particular choice of the knots, $\xi_i$,  distribution and the smoothing penalty coefficient\footnote{See section \ref{DLI_III_sect_stat_anal} for the definition of the smoothing penalty.}, $q$. Therefore, each set of parameters $\{\theta_{\star}, \theta_{\bullet}, \boldsymbol{\xi}, q \}$ defines a unique model.

Using a B-spline approximation for the radial velocity dispersion, $\sigma_{\text{rr}}^2$, we avoid any bias from a specific assumption on the functional form of the anisotropy profile. Effectively, we transform the set of all possible choices for the functional form of the anisotropy profile, $\beta(r)$, from a countable infinite discrete set, to an infinite yet continuous set of $\sigma_{\text{rr}}^2$ 
solutions. 
This continuity in the variation of the possible kinematic profiles  significantly limits the number of models that we need to compare in order to choose the best one that closest approximates  reality. The total number of models now is restricted to the total number of stellar tracer and DM model combinations.

\subsubsection{The need for  adaptive knot distribution in \jeans modelling}

The B-spline approximation of a function, Eq. \eqref{DLI_III_gen_func},  depends on the quality of the data we have and the total number and geometric shape of the basis functions, $B_i(x)$.  The latter are defined from the choice of the knot vector, $\boldsymbol{\xi}$, and their order, $k$. A large knot vector results in a large number of basis functions, and therefore a large number of unknown coefficients $a_i$. This increases the dimensionality of the unknown parameters, and may result in overfitting. On the other hand a bad choice of knots, $\xi_i$, can result in a bad approximation of the function - even if this is only locally. The problem of finding the optimum choice of knots, $\xi_i$, is a nonlinear one and is generally  very difficult to tackle. This is why in the  literature the proposed method is to use a large number of uniform knots; then one regulates over-fitting or under-fitting with the use of a single smoothing parameter that is related to the penalty of the curvature of the fitted curve. See \cite{hastie01statisticallearning}  and references therein for a description of the problem.

In order to get the best function approximation for $\sigma_{\mathrm{rr}}^2$ and $\sigma_{\mathrm{los}}^2$, with the least number of unknown coefficients, we need b-splines that are locally adaptive. This cannot be solely through the optimum smoothing parameter because it is a global regulator  and cannot account for local irregularities in the data. This is more apparent when our data are incomplete with large errors (as is usually the case with line-of-sight velocity dispersion, $\sigma_{\text{los}}^2$, values), or 
the function to be approximated has parts with a steep slope (as is the case for many of the mock suites of the \textsc{GAIA Challenge}\footnote{ Wiki site http://astrowiki.ph.surrey.ac.uk/dokuwiki/.   The performance of the \jeans is currently tested with the   \textsc{Gaia Challenge} mock suite in Diakogiannis et al. in prep.} data sets).

One important property of B-spline functions, in the context of Galactic Dynamics, is their local support\footnote{
For a B-spline function of order $k$, $f(x) = \sum_i a_i B_i(x)$,   at each location $x \in [\xi_i, \xi_{i+1}) $ there are at most $k$ non zero basis functions, namely the $\{ B_{i-k+1}(x), \ldots, B_i(x) \}$. This means that the corresponding coefficients that determine the geometric shape of the function at $x$ will be the $\{ a_{i-k+1},\ldots, a_i \}$.}. In practice, this means that we can make local changes in the geometric shape of a curve without affecting its  geometric shape globally. 
For example, when  we fit a smoothing B-spline curve to data, the value of the curve at a given location, $x$, depends on the data in the local region around $x$. 
How ``local'' the region around $x$ is depends crucially on the knot vector, $\boldsymbol{\xi}$, that defines the basis $B_i(x)$. 
In the \jeans formalism the $\sigma_{\mathrm{los}}^2(R)$ profile is determined by two factors: the data, and the smoothing penalty.   The local support of the B-spline functions is inherited in the $\sigma_{\mathrm{los}}^2$ profile as well. Since our data extend up to a limited distance from the origin\footnote{In fact the virial radius of a system extends much further than the location of the last datum.} of the system, we want to have adaptive splines such that the unknown coefficients $a_i$ are \textit{all} influenced by the data. In the extreme case where we have a very large number of unknowns, the value of the coefficients, $a_i$, that correspond to locations beyond the last datum, will be solely determined by the smoothing penalty and result in a straight line.

Furthermore,  when we want to compare two DM models, we would like to have kinematic profiles that are described in the least complex way. All the above make the evaluation of the optimum knot distribution an important part of our work.  See Appendix \ref{DLI_III_simple_curve_Example} for a simple example on the significance and impact of the EA-best knot distribution chosen by our algorithm versus a uniform $\boldsymbol{\xi}$ for  simple curve fitting.

\subsection{Statistical Analysis}
\label{DLI_III_sect_stat_anal_primary}

In this section we  describe our likelihood model, as well as the process for obtaining marginalized distributions of our model parameters. See  \cite{2014MNRAS.443..610D} for an analytical justification of our choices of the various functional forms. 
 
The general form of the posterior probability we use is:
\begin{equation}\label{DLI_III_loglkhood}
\mathcal{L}=
p(\theta|I)\; p(\tilde{q}|\alpha,\beta)\;  p(W|\tilde{q}) \;  L(D|\theta,I)
\end{equation}
For a fixed knot distribution, $\boldsymbol{\xi}$, $p(\theta|I)$ is the prior range of all model parameters $\theta=\{\theta_{\star}, \theta_{\bullet}, \Upsilon_V, a_i, q \}$, which we take to be uninformative (uniform). $p(\tilde{q}|\alpha,\beta)$ is the hyperprior on the smoothing parameter, $\tilde{q}$.  $p(W|\tilde{q})$ is the smoothing penalty term and $L(D|\theta,I)$ is the likelihood of the model. In full detail these equations are: 
\begin{align}
\label{DLI_III_HyperPriorparams}
p(\tilde{q}|\alpha,\beta) &= 
\frac{\beta^{\alpha}}{\Gamma(\alpha)}
\tilde{q}^{-\alpha-1} e^{-\beta/\tilde{q}}
\\
\label{DLI_III_W}
W &=\int_{R=0}^{r_{\text{tt}}}\left( \frac{\mathrm{d}^2 \sigma_{\text{los}}^2}{\mathrm{d} R^2}\right)^2 \mathrm{d} R\\
p(W|\tilde{q}) &= \tilde{q} \exp(-\tilde{q} W)\\
\label{DLI_III_q_trans}
\tilde{q} &= \frac{1-q}{q}, \quad q\in (0,1]
\end{align}
and
\begin{multline} \label{DLI_III_lkhood_fll}
L(D|\theta,I) = \prod_{i=1}^{N_{\star}} 
\exp\left[ -\frac{(\Sigma_i-\Sigma_{\star}(R_i))^2}{2 (\delta \Sigma_i)^2}\right] / \sqrt{2 \pi (\delta \Sigma_i)^2} \\ \times 
\exp\left[ -\frac{(\sigma^2_i-\sigma_{\text{los}}^2(R_i) )^2}{2 (\delta \sigma_i)^2}\right] / \sqrt{2 \pi (\delta \sigma^2_i)^2} 
\end{multline}
In Eq \eqref{DLI_III_lkhood_fll} the variable $\Sigma_i$ is the surface density  at position $R_i$, and $\delta \Sigma_i$ its uncertainty. Similarly, $\sigma^2_i$ is the observed line-of-sight velocity dispersion, and $\delta \sigma^2_i$ the uncertainty. We need to stress that the line-of-sight velocity distribution function (LOSVDF) is not Gaussian \citep{1987ApJ...313..121M}. 
In addition even if the LOSVDF were Gaussian the distribution function for the $\sigma_{\text{los}}^2$ would follow a $\chi^2$ distribution which has much wider tails. 
Our likelihood in Eq \eqref{DLI_III_lkhood_fll} should be viewed as the Bayesian approach of the weighted $\chi^2$ penalty function under the assumption of  uncorrelated weights:
\[
\chi^2 \sim  \sum_i \left( \frac{\sigma_i^2 - \sigma_{\text{los}}^2(R_i)}{\delta \sigma_i} 
\right)^2
\]
The quantity $W$ is the smoothing penalty factor\footnote{In this contribution we are using only the second order derivative of $\sigma_{\text{los}}^2$. In \cite{2014MNRAS.443..610D}  we have used both first and second order derivatives. The main reason for our choice is computational efficiency.}, while the function $\tilde{q}(q)$ is a convenient way of restricting the smoothing penalty parameter range to $(0,1] \ni q$ for computational purposes. 
 The prior on the $\tilde{q}$ values,  $p(\tilde{q}|\alpha,\beta)$, is the inverse Gamma distribution\footnote{See \cite{2014MNRAS.443..610D} for the justification of this choice of hyperprior.}, and the values of parameters $\alpha, \beta$ will be chosen from mock  datasets of stellar systems in a process described later. This is a crucial part of our algorithm: we use information from theoretical models to identify the optimum smoothing value range for $\tilde{q}$. This is an extension of Empirical Bayes methods \citep{1985} that defines prior information from moments of the data: here we encode some specific characteristic of ideal theoretical models (specifically the optimum smoothing of $\sigma_{\text{rr}}^2$) in order  to build prior information.

\section{The JEAnS Algorithm}
\label{DLI_III_AlgoOverview}

\begin{figure}
\centering
%height=0.3\textwidth, width=0.5\textwidth
%\showthe\columnwidth 
\includegraphics[width=\columnwidth]{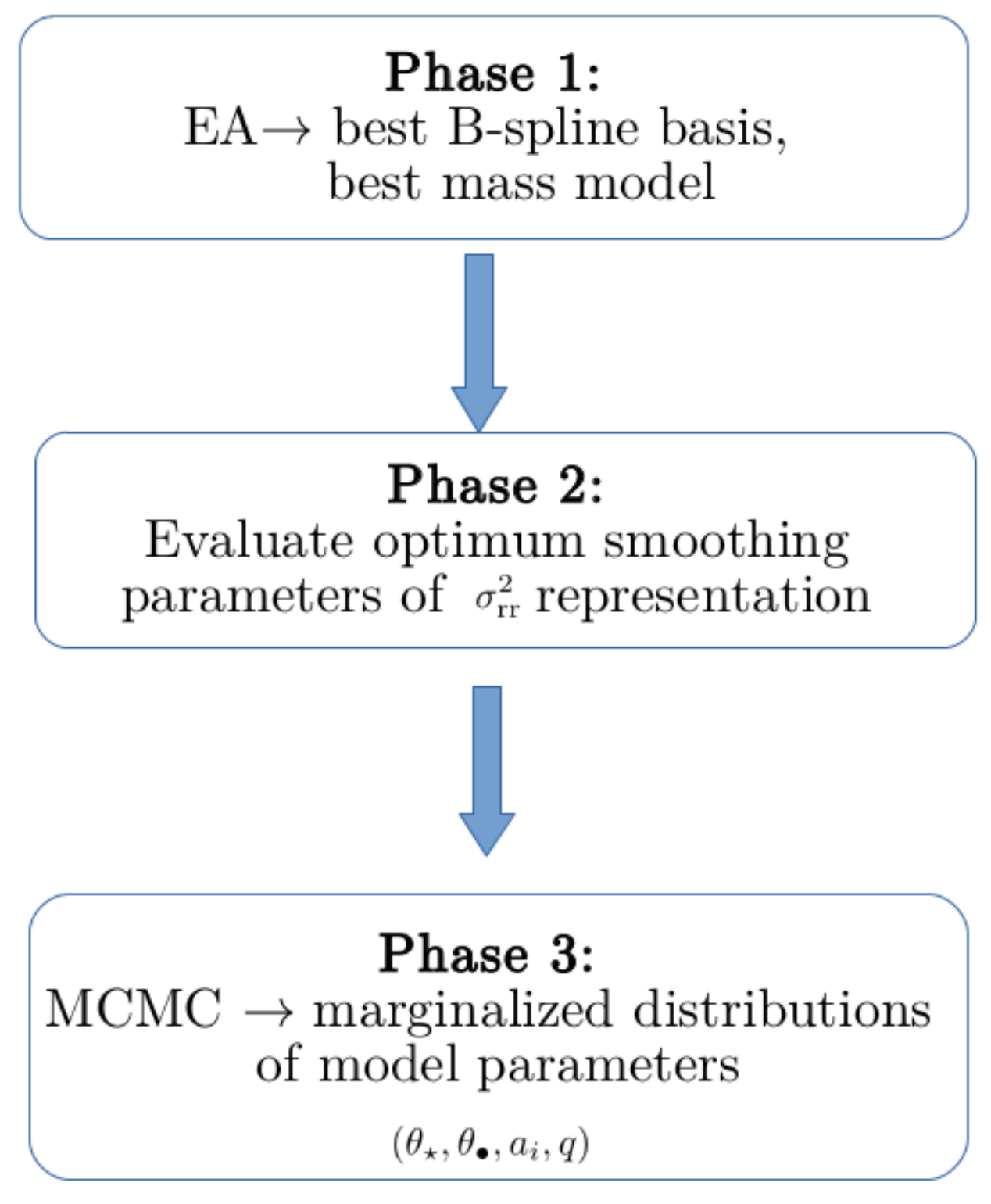}
\caption{Flow diagram of the {\sc JEAnS} algorithm. It consists of 3 phases. In phase 1, the EA is used for the evaluation of optimum knot distribution and best mass model. In phase 2 we use information from ideal theoretical models in order to determine the optimum smoothing parameters. In phase 3 we use MCMC to estimate the marginalized distributions of the model parameters, keeping the knots fixed.}
\label{DLI_III_FlowChart}
\end{figure}

In this section we describe the steps of the {\sc JEAnS} algorithm.  
The method we propose is a three (3) phase method.  The three phases are the following:  
\begin{enumerate}
\item In phase 1 (section \ref{DLI_III_EA_model_selec_sect}), we evaluate the optimum  knots, $\{\xi_i \}$, and discriminate between competing mass models.  
\item In phase 2 (section \ref{DLI_III_OptSmoothing_sec}) we encode information for the optimum smoothing of the B-splines from  ideal theoretical models.  
\item  In phase 3 (section \ref{DLI_III_sect_stat_anal}) we have fixed knots and smoothing parameters (selected appropriately from phases 1 and 2) and perform statistical regression for the 
model parameters via Markov chain Monte Carlo (MCMC). 
\end{enumerate}
We represent schematically the flow diagram of  this algorithm in Fig. \ref{DLI_III_FlowChart}.

\subsection{Evolutionary model selection}
\label{DLI_III_EA_model_selec_sect}

In this section we will state the nature of the modelling problem and how our algorithm tackles it. Then we will describe the framework and the particular characteristics of the evolutionary algorithm we developed. For the convenience of the readers we try to avoid as much as possible the standard terminology from the EA literature. In Appendix 
\ref{DLI_III_EA_glossary} 
we give  a small glossary of terms. Readers who wish to understand more about evolutionary algorithms can consult  \cite{Goldberg:1989:GAS:534133} for a first understanding and \cite{Talbi:2009:MDI:1718024, Rozenberg:2011:HNC:2016676} for the state of the art in evolutionary computation.

\subsubsection{The need for evolutionary algorithms}

By approximating the radial velocity dispersion, $\sigma_{\text{rr}}^2$, with a B-spline function, we introduce a variable number of unknown variables: the knot vector $\boldsymbol{\xi}$ as well as the unknown coefficients $a_i$. So the problem of fitting our model to a particular data set now becomes: find the best set of parameters for the stellar model, $\theta_{\star}$, the DM model, $\theta_{\bullet}$, the mass-to-light ratio, $\Upsilon_V$, the (variable length) knot vector, $\boldsymbol{\xi}$, and the (variable number of) unknown coefficients, $a_i$, of $\sigma_{\text{rr}}^2$. This is  a very computationally demanding problem 
 mainly due to the variable number of parameters. 
 
 EAs have the advantage that they are robust in finding a global optimum solution as well as being very fast. They also do not require the fitness function to be a probability distribution function, so any particular condition we consider as physical (that is not easy to be described in terms of some probability density function) can be encoded. Furthermore, they can easily treat variable number of unknowns. Thus by using as a fitness function a model selection criterion, like the Akaike Information Criterion \citep[AICc, ][]{sugiura1978further, hurvich1991bias}, we can effectively fit and select the best model at each iteration.

\subsubsection{Solution representation}

In this section we  describe how we represent the solution inside the EA. 
Each candidate solution (termed individual in the EA framework) consists of two parts: the first is a  binary string of variable length  that represents the variable length knot vector,  $\boldsymbol{\xi}$. The second is a vector of  real variables, $\theta$, of fixed length that represents the stellar and dark matter model parameters, as well as the smoothing penalty coefficient $q$ and constant mass-to-light ratio, $\Upsilon_V$. Therefore, an individual is represented as a tensor product of the two different types of variables,  $\xi_i \otimes \theta_j$.

 The binary string that represents the knot vector, $\boldsymbol{\xi}$, has length $N_{\text{knots}} \times N_{\text{bits}}$, where $N_{\text{knots}}$ is the total number of individual knots, and $N_{\text{bits}}$ is the length of the binary string that represents a single knot, $\xi_i$.  For  $N_{\text{bits}}=6$ 
bits\footnote{The first and last knots are fixed to real values of $0$ and $1$, i.e. $\xi_1 = 000000$, $\xi_n=111111$.} 
we have  127 subdivisions of the real interval $[0,r_{\text{t}}]$. This is a very good approximation for the quality of data we have in hand. The large uncertainties of the observables make the need for a further refinement unnecessary. 
Internally each of the knots, $\xi_i$, is evaluated in the interval $[0,1]$. Then the  vector of knots is multiplied by the tidal radius of the system, as estimated from the King model, and thus the knots take positions in the real space of variables. In the case where a different stellar profile is used, we  define the tidal radius as the virial radius of the DM halo.

\subsubsection{Recombination}
In this section we  detail the process of \textit{recombination}, i.e. the combination of solutions from the EA population between individuals that produce new solutions (offspring). Recombination consists of the operation of 
\textit{crossover}, where individuals exchange ``genetic material'' (they combine their representation values), and \textit{mutation}, where a single individual has some of its variables randomly changed.

As mentioned previously, each individual representation consists of two types of variables, a binary string for the representation of knots, $\xi_i$, and a vector of reals for all the model parameters, $\theta$, i.e. $\xi_i \otimes \theta_j$. 
Recombination takes place between variables of the same type. For the real coded variables, ${\theta}$, the crossover we use is  PCX \citep{Deb:2002:CEE:638598.638601}, and the mutation operator is the Makinen, Periaux and Toivanen Mutation (MPTM)  operator \citep{FLD:FLD829}. For the binary string of knots we had to devise our own crossover and mutation operators. We therefore detail below the operation of recombination only for the knot vectors $\boldsymbol{\xi}$ between two individuals.

\subsubsection{Knot crossover}

\begin{figure*}
\centering
%height=0.3\textwidth, width=0.5\textwidth
%\showthe\columnwidth 
\includegraphics[width=\textwidth]{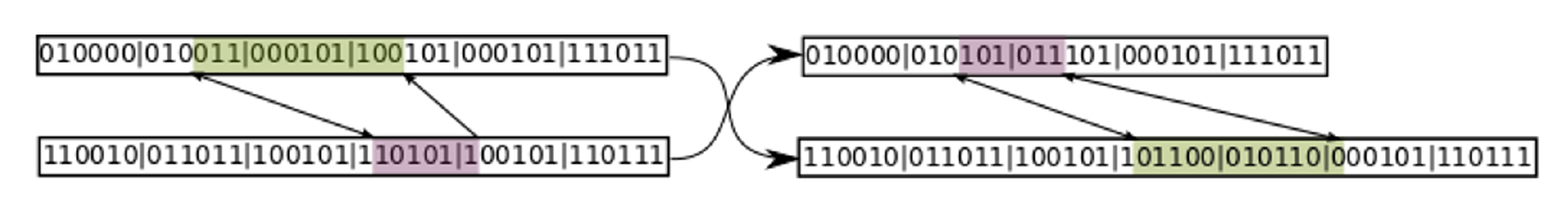}
\caption{Crossover operation between the knots, $\boldsymbol{\xi}$,  of two individuals. For each individual a region is selected at random that corresponds to an integer multiple of the number of bits that can represent a knot. Then these two regions are exchanged. The number of knots for each individual is not preserved, however the total number of knots between both individuals is preserved. }
\label{DLI_PR_knot_crossover}
\end{figure*}

We want the crossover operation between two knots to have the following characteristics: combine information from the total  number and the value of the knots for each individual. Therefore, we perform a modified version of the two point binary crossover: 
We select two segments of binary strings from each individual, such that the total length of each segment corresponds to an integer multiple of a single knot, $\xi_i$, of  length $N_{\text{bits}}$. Then these segments are exchanged and the resulting binary strings are sorted in ascending order.  For the case where the size of the recombined individuals is 3\footnote{Recall that the first and last knots are held fixed to $\xi_1=0$, $\xi_m=1$.}, i.e. $\boldsymbol{\xi} =(\xi_1,\xi_2,\xi_3) = (0,\xi_2,1)$,  the crossover is a standard two point crossover between the middle knots with the restriction the length of the substrings exchanged  to be equal.  With these operations the total number of knots remains invariant, i.e. the crossover operation cannot modify the total number of knots.  
The crossover operation between two knots is summarized in Algorithm \ref{DLI_II_Algo_cross}. A schematic representation of the knot crossover is presented in Fig. \ref{DLI_PR_knot_crossover}.

\begin{algorithm}[H]
\begin{algorithmic}[1]
\Require{Parents $\chi_1, \chi_2$}
\Ensure{Offspring $\tilde{\chi}$}
\If{$\dim_1=\dim_2=3$}
\State $(\tilde{\chi}_1,\tilde{\chi}_2) \gets (\chi_1,\chi_2)$ 
\Comment{standard two point crossover}
\Function{Repair}{$\tilde{\chi}_1$}
\State {\bf Sort} $(\tilde{\chi}_1)$ \Comment{$\xi_i,\xi_j \in (\tilde{\chi}_1)$,   $\xi_i\leq \xi_j$ for $i\leq j$}
\EndFunction
\Function{Repair}{$\tilde{\chi}_2$}
\State {\bf Sort} $(\tilde{\chi}_2)$
\EndFunction
\Else
\State Calculate random $n_1,n_2\in [1,N_1]$
\State Set $\{n_1,n_2\}=\{\min(n_1,n_2),\max(n_1,n_2)\}$
\State Calculate integer $a=(n_2-n_1)/N_{\text{bits}}$
\State Set ($n_2=a n_1$)
\Comment{Ensures multiple of $N_{\text{bits}}$}
\Function{Copy}{$(n_1,n_2)$}
\State Copy bits from $\chi_1$ between $[n_1,n_2]$
\EndFunction
\State Repeat the same for $\chi_2$
\State  $(\tilde{\chi}_1,\tilde{\chi}_2) \gets (\chi_1,\chi_2)$ 
\Comment{Exchange binary segments}
\Function{Repair}{$\tilde{\chi}_1$}
\State {\bf Sort} $(\tilde{\chi}_1)$  
\EndFunction
\Function{Repair}{$\tilde{\chi}_2$}
\State {\bf Sort} $(\tilde{\chi}_2)$ 
\EndFunction
\State $\tilde{\chi} \gets$ best$(\tilde{\chi}_1,\tilde{\chi}_2)$
\EndIf
\end{algorithmic}
\caption{Crossover operator for knots $\boldsymbol{\xi}$}
\label{DLI_II_Algo_cross}
\end{algorithm}

\subsubsection{Knot mutation}

In order to get diversity in the knots, we want the mutation to be able to change both the number of knots as well as their value. Therefore, we split the mutation operator in two operations, each taking place with probability $p=0.5$: half of the times the mutation will only modify the number of knots, the other half only their value. When we modify the number of knots, we decide if we will delete, or add a random knot according to a biased Bernoulli probability, $\text{flip}(p_{\mathrm{m}})$, where $p_{\mathrm{m}}=0.005$. If $\text{flip}(p_{\mathrm{m}})=true$ we either add or delete a knot, at a random position, with probability $p=0.5$.  When we want to modify the value of knots, we perform a standard binary flip bit mutation as described in \cite{Goldberg:1989:GAS:534133}, with probability $\text{flip}(p_{\mathrm{m}})$. In this way the mutation operator preserves diversity in both the number of knots as well as their value. We summarize the process in Algorithm \ref{DLI_III_mutation}. The variable $N_{\text{bits}}$ corresponds to the total number of binary digits of the knot vector, excluding the first and last knots, which remain fixed to 0 and 1 respectively.

\begin{algorithm}
\begin{algorithmic}[1]
\Require{$\chi$  Chromosome}
\Ensure{$\tilde{\chi}$ Mutated individual}
\State Calculate random $p_1=\text{unif}(0.1)$
\If{$p1 \leq 0.5$} 
\Comment{Flip bits}
\For{$i=1\ldots,N_{\text{bits}}$}
\If {$ \text{flip}(p_{\mathrm{m}}):$ true}
\Comment{Bernoulli biased} 
\State Flip bit
\EndIf
\EndFor
\Else
\Comment{Modify  number of knots}
\If {$\text{flip}(p_{\mathrm{m}})$}
\Comment {Add/delete  random knot}
\State Calculate $p_2=\text{unif}(0,1)$
\If {$p_2 \leq 0.5$}
\State Insert a random bits knot
\Else 
\State Delete a knot
\EndIf
\EndIf
\EndIf
\Function{Repair}{$\tilde{\chi}$}
\State {\bf Sort} $(\tilde{\chi})$ \Comment{$\xi_i,\xi_j \in (\tilde{\chi})$,   $\xi_i\leq \xi_j$ for $i\leq j$}
\EndFunction
\end{algorithmic}
\caption{Mutation operator}
\label{DLI_III_mutation}
\end{algorithm}

\subsubsection{Evolution model}
\label{DLI_III_EvoMod}

For the evolution of the EA, we used a modified version of the Generalized Generation Gap evolutionary model \citep[][hereafter G3]{Deb:2002:CEE:638598.638601}. 
The modification is necessary in order to account for the two distinct types of variables in the individual representation, namely the variable length binary string, $\boldsymbol{\xi}$, and the vector of reals, $\theta$. We chose using the G3 evolutionary model because of its competitive performance over other evolution schemes \citep{Hansen:2010:CRA:1830761.1830790}.

In the G3 algorithm, we create a temporary  pool, $P_{\mu-1}$,   of parents  consisting of  $\mu-1$ randomly chosen individuals. Each of the $\mu-1$ parents is selected with equal probability from the population $P$, excluding the best candidate, $\hat{\chi}$. Then we create $\lambda$ offspring with the following scheme: 
\begin{enumerate}
\item Select a random individual, $\chi_i$, from the temporary pool, $P_{\mu-1}$, with uniform probability.  
\item Crossover the knot vector, $\hat{\boldsymbol{\xi}}$, of the best individual, $\hat{\chi}$ with the knot vector, $\boldsymbol{\xi}_i$, of  $\chi_i$ (this operation produces a single knot vector). 
\item Crossover the vector of reals, $\theta$ of the best individual,  $\hat{\chi}$, with the vector of reals from the pool $P_{\mu-1}$, according to the PCX crossover scheme \citep{Deb:2002:CEE:638598.638601}.  This operation uses the information from the whole pool of parents, $P_{\mu-1}$, and produces one vector of reals centered close to the best candidate solution. 
\item Mutate knots.  
\item Mutate $\theta$ according to the MPTM algorithm. 
\item The offspring consists of the new knot vector and the new generated vector $\theta$.  
\end{enumerate}
This process is repeated until $\lambda$ offspring are created. Then we select two candidates, $\chi_1$ and $\chi_2$,  from the initial population, $P$. We construct an augmented pool, $P_{\lambda+2}$, that consists of $\lambda$ offspring, $P_{\lambda}$,  and the $\chi_1$ and $\chi_2$. The two best candidates of this augmented pool, $P_{\lambda+2}$, replace the initial $\chi_1$ and $\chi_2$ in the  population, $P$. The modified G3 algorithm is summarized in Algorithm \eqref{Algo_G3}. This evolution method preserves the best candidate from the previous population if it remains the best solution after the recombination operation.

The algorithm set up is: total number of individuals in the population $N_{\text{pop}}=100$, number of parents $\mu = 5$, number of offspring $\lambda = 20$. 
We run the algorithm for a sufficient number of generations until convergence (usually a few thousand generations). 
For the MPTM mutation we used a constant probability of $p_{\mathrm{m}}=0.005$. Unlike traditional genetic algorithms, in G3 the crossover operation  is always performed (crossover probability $p_{\mathrm{c}}=1$) and by design it can jointly explore all parameter space and maintain diversity. We added mutation for additional diversity in the solution space.

% Algorithmicx version of G3: 
%\begin{comment}
\begin{algorithm}
\begin{algorithmic}[1]
\Require {Population $P$}
\Ensure {Updated population $P$}
\Repeat
\State Select the best individual, $\tilde{\chi}$.
\State Select  $\mu-1$ random individuals from $P$. 
\State $P_{\mu-1}=\{ \chi_j\}$, $j=1,\ldots,\mu-1$
\Procedure {Create $\lambda$ offspring}{$\tilde{\chi},P_{\mu-1}$}
\State $P_{\lambda}\gets \{\}$ \Comment{Empty pool}
\For{$i=1,\ldots,\lambda$ offspring}
\State random $\chi_i \in P_{\mu-1}$. 
\State $\chi_i \gets$  \Call{Crossover}{$\tilde{\chi},\chi_i,P_{\mu-1}$}
\State $\chi_i \gets$ \Call{Mutate}{$\chi_i$}
\State $P_{\lambda}\gets \chi_i$
\EndFor
\State {\bf return} $P_{\lambda}$  
\EndProcedure
\State Select  random $\chi_1,\chi_2 \in P$.
\State $P_{\lambda+2} \gets \{\chi_1,\chi_2\}\cup P_{\lambda}$
\State $ P \ni(\chi_1,\chi_2)  \gets  (\hat{\chi}_1,\hat{\chi}_2)  \in (P_{\lambda+2})$.
\Until Convergence
\end{algorithmic}
\caption{G3 evolution model}
\label{Algo_G3}
\end{algorithm}
%\end{comment}

\subsubsection{Fitness function}
\label{DLI_III_fitness_section}

The  unknown parameters that fully specify a solution,  are the knots, $\xi_i$, the coefficients $a_i$, the stellar mass density parameters, $\theta_{\star}$, the DM mass density parameters, $\theta_{\bullet}$, the mass-to-light ratio, $\Upsilon_V$, and the smoothing penalty coefficient, $q$.  
Let $\theta =\{\xi_i,\theta_{\star},\theta_{\bullet},\Upsilon_V  \}$ denote the subset of these that does not contain the unknown coefficients, $a_i$, and smothing parameter, $\tilde{q}$. 
Due to the linearity of the equation \eqref{DLI_III_sigmaLOS} with respect to the $a_i$ coefficients, once we specify a set of values $\theta$ and $\tilde{q}$  we can find a unique solution for the set of unknowns, $a_i$. 
This follows from the form of the $\chi^2$ minimization of the penalty function to obtain the $a_i$:  
\begin{equation}
\label{DLI_III_QP}
-\ln\mathcal{L}_{a_i}=
\sum_{i=1}^{N_{\text{data}}}
\left[ \frac{(\sigma^2_i-\sigma_{\text{los}}^2(R_i) )}{ \delta \sigma^2_i} \right]^2
+ \tilde{q}W
\end{equation}
where $\sigma^2_i$ are the data values of the line-of-sight velocity dispersion, $\sigma_{\text{los}}^2$; $\tilde{q}$ and $W$ are given by  Eq. \eqref{DLI_III_q_trans} and Eq. \eqref{DLI_III_W} (detailed in section \ref{DLI_III_sect_stat_anal}).
The functional form of   $\ln\mathcal{L}_{a_i}$ is of convex type with respect to the $a_i$ unknowns\footnote{Least squares minimization with a smoothing penalty is equivalent to quadratic programming minimization.}, hence it has a unique solution. 

Therefore, we break down the optimization problem to a hybrid one: a combination of heuristic search  for the $\theta$ and $\tilde{q}$ parameters that are being treated by the EA, and a convex minimization for the evaluation of the $a_i$. The convex minimization takes place inside the EA\footnote{In practice we also evaluate $\tilde{q}$ inside the EA, since specifying $\theta$ yields a unique solution for $a^i$ and $\tilde{q}$. This problem, however, is non-convex.} at each $\ln\mathcal{L}_{a_i}$ evaluation. In this way we significantly speed up the convergence of the overall optimization. Furthermore, using convex minimization algorithms, we can easily encode meaningful restrictions to the unknown B-spline coefficients, $a_i$, in order to avoid non physical solutions such as\footnote{The restriction $a_i \geq 0$ we used in \cite{2014MNRAS.443..598D,2014MNRAS.443..610D} is a stronger requirement than $\sigma_{\text{rr}}^2 \geq 0$ but it cannot guarantee that $\sigma_{\text{los}}^2 \geq 0$.} $\sigma_{\text{rr}}^2, \sigma_{\text{los}}^2 \geq 0$. 
For the convex minimization we used quadratic programming routines, specifically the software qpOASES\footnote{Source:  http://www.qpOASES.org/}  \citep{Ferreau2014}.

For a maximization problem, the  fitness function is related to the transform of the AICc (corrected Akaike Information Criterion): 
\begin{equation}
\label{DLI_III_fitFunction_general}
f(\theta) = \frac{1}{1+\text{AICc}(\theta)}
\end{equation}
where
\begin{equation}
\label{DLI_III_AICc_general}
\text{AICc} = -2 \ln \mathcal{\hat{L}}_{\text{EA}} + 2 n + \frac{2n (n+1)}{N_{\text{data}}-n-1}
\end{equation}
and $n$ is the number of unknown parameters. 
AICc is the Akaike Information Criterion (AIC) with second order bias correction  \citep[see][and references there in]{opac-b1100695},   
and
\begin{multline}
\label{DLI_III_logLEA_trainError}
-\ln \mathcal{L}_{\text{EA}} = \sum_{i=1}^{N_{\text{data}}}
\left[ \frac{(\sigma^2_i-\sigma_{\text{los}}^2(R_i) )}{ \delta \sigma^2_i} \right]^2
\\ +\sum_{i=1}^{N_{\text{data}}} 
\left[  \frac{(\Sigma_i-\Sigma(R_i))}{\delta \Sigma_i} \right]^2 
\end{multline} 
AICc is the approximation of AIC for a small number of data compared to the number of unknowns.  We used AICc instead of BIC  since the former is more robust for small datasets while the latter is inappropriate when the number of parameters is comparable to the number of available data \citep{opac-b1100695}. AICc is  known to give the best model and solution according to the optimum trade off between bias and variance \citep{hastie01statisticallearning}. 

When we evolve the EA we use $n=n_{\text{coeffs}}$ in Eq. \eqref{DLI_III_AICc_general}, i.e. $n$ is the total number of unknown coefficients, $a_i$, of the B-spline approximation $\sigma_{\text{rr}}^2$, not the total number of model parameters, $n_{\text{tot}}$. This stems from the fact that we are interested to identify the smallest possible knot vector that gives a good  fit.  
Once we have established a good fit and  we want to compare  models, we calculate the AICc using all the model parameters. The reason we do this is related to multiobjective optimization 
\citep[][see also \citealt{ghosh2008multi}]{journals/swevo/ZhouQLZSZ11}: the AICc has two goals, to minimize the number of unknowns, $n$, and find the best fit by minimizing $-2 \ln \hat{\mathcal{L}}_{\text{EA}}$. So in principle we have a multiobjective (i.e. vector) optimization problem of the form, $\mathbf{f} = ( -2 \ln \hat{\mathcal{L}}_{\text{EA}}, n )$ of two conflicting goals, where $\mathbf{f}$ is a two dimensional fitness function. Combining both information criteria into one scalar function, i.e. the AICc,  gives just one solution instead of the whole Pareto front \citep{Coello:2006:EAS:1215640} of equally good solutions. This also means that the optimization is sensitive to the relative values of the terms $\ln \hat{\mathcal{L}}_{\text{EA}}$ and $n$: if we increase the number of parameters, $n$, in the AICc by using the total number of parameters, $n_{\text{tot}}$,  we bias the solution towards smaller number of knots. This is especially more profound when we have a small number of data and $\ln \hat{\mathcal{L}}_{\text{EA}}$ is in the same range of values with $n$. 
After experimenting with various synthetic data sets we conclude that using in the AICc only the unknown coefficients, $n_{\text{coeffs}}$, for the EA fit, yields the best kinematic profile. 
 After the EA optimization, by using the total number of parameters $n_{\text{params}}$ in the AICc, we can perform model inference between competing mass models.

\subsection{Evaluation of optimum smoothing parameters}
\label{DLI_III_OptSmoothing_sec}

In this section we describe how we calculate the value of the parameters $\alpha, \beta$ of the hyperprior $p(\tilde{q}|\alpha,\beta)$ in order to evaluate uncertainties in all of the model parameters through a Markov Chain Monte Carlo process. 
The method we propose here is significantly different than what we presented in \cite{2014MNRAS.443..610D}.

The hyperprior distribution $p(\tilde{q}|\alpha,\beta)$ regulates the values selected for the $q$ parameter from the MCMC algorithm that fits the model (now with fixed knots) to the data. Then we need a  set of $\{q_j\}$ data values in order to fit the $p(\tilde{q}|\alpha,\beta)$ distribution to these, and obtain estimates for the $(\alpha,\beta)$ parameters. 
\begin{comment}
These $\{q_j\}$ data values will be estimated from  
mock stellar systems created  from  ideal theoretical functions, e.g. 
 that  approximate the system we are investigating.  
\end{comment}

Having evaluated a set of optimum  model parameters, $\theta$ with the use of the EA, we can use these as reference  parameters (hereafter $\theta^{\text{ref}}$) to calculate synthetic data from some reference anisotropy profile, $\beta^{\text{ref}}(r)$. We are free to choose any form of the anisotropy profile but we would like to use a profile that has similarities with the best knot vector solution from the EA. For example we can use the generalized  Osipkov-Merritt 
\citep{1979SvAL....5...42O,1985AJ.....90.1027M}
anisotropy profile
\begin{equation}
\label{DLI_III_OM}
 \beta(r)=c_1 \frac{r^2}{r^2+c_0^2}
\end{equation}
with $c_0$ equal to the mean value of the interior knots. Alternatively we can also fit the system under investigation with the traditional method of Jeans modelling  and use the highest likelihood fit as an ``ideal theoretical model'', i.e. $\theta^{\text{ref}},\beta^{\text{ref}}$,  from which we will draw conclusions for the smoothing hyperprior parameters, $(\alpha, \beta)$.

For this reference profile, $(\theta^{\text{ref}},\beta^{\text{ref}})$, we can estimate a reference B-spline approximation\footnote{We can do so, by solving the equation 
\[
\sigma_{\text{rr}}^2|_{\text{ref}}(r) = \sum_i a_i^{\text{ref}}B_i(r)
\]for a large number of data values, $\{(r_i,\sigma_{\text{rr}}^2(r_i)|_{\text{ref}}) \}$, with no smoothing penalty and standard linear algebra operations. We recommend linear or quadratic programming solvers, since these can incorporate physically plausible constrains, e.g. $\sigma_{\text{rr}}^2 \geq 0$.} of the radial velocity dispersion, $\sigma_{\text{rr}}^2|_{\text{ref}}(r) = \sum_i a_i^{\text{ref}}B_i(r)$. 
The best smoothing parameter, $q$, is the one whereby using $\theta_{\text{ref}}$ as model parameters, and $q$ as  the smoothing parameter, the estimated solution, $\hat{a}_i$, has the closest proximity to the reference values, $a_i^{\text{ref}}$.   
We use the following measure for the optimum penalty parameter $q$: 
\begin{equation}
\label{DLI_III_smoothing_pnlty}
S(q) = \sum_{j=0}^2 \sum_{i=1}^n | \Delta^j \hat{a}_i(q)  -
 \Delta^j a^{\text{ref}}_i |,
\end{equation} 
where $\hat{a}_i(q)$ emphasizes the dependence of the solution values $\hat{a}_i$ on the smoothing parameter, $q$.  
We estimate the coefficients $\hat{a}_i$ using the quadratic programming minimization of Eq. \eqref{DLI_III_QP}. 
  The ``differences''
\begin{align*}
\Delta^0 a_i &=  a_i\\
\Delta^1 a_i &=  a_{i+1} - a_i\\
\Delta^2 a_i &=  a_{i+2} - 2 a_{i+1} +  a_i
\end{align*}
represent respectively: the difference in the coefficients value, $\Delta^0$, the  approximate slope, $\Delta^1$, and the  approximate curvature, $\Delta^2$. That is, we require $C^0$, $C^1$ and $C^2$ proximity of the estimated solution to the reference profile. 
To get a robust set of values $q$ that minimize Eq. \eqref{DLI_III_smoothing_pnlty}, we  create random realizations of synthetic data $(R_j,\sigma_{\text{los}}^2(R_j))$, based on the reference profile. These random data sets have the same positions $R_j$ as the true data set we wish to fit, and are equal in number\footnote{See \cite{2014MNRAS.443..610D} for more details.}. 
Then for each of these we minimize the function $S(q)$, with respect to the $q$ parameter, using some black-box non-linear optimizer and obtain a value $q_j$. Our choice of  robust blackbox   optimizer is the method Covariance Matrix Adaptation Evolutionary Strategy  \citep[CMA-ES][]{Hansen96adaptingarbitrary}, and the {\sc C++} software implementation  {\sc libcmaes}\footnote{Source: https://github.com/beniz/libcmaes}. We summarize this process in Algorithm  \ref{DLI_III_Opt_smooth}. 
Once we have a set of $\{ q_j \}$ data values, we fit the hyperprior function $p(\tilde{q}|\alpha, \beta)$ to these and get estimates for the  parameters $(\alpha, \beta)$.

\begin{algorithm}[H]
\begin{algorithmic}[1]
\Require{Ideal model $\bar{\theta}=\{ \theta^{\text{ref}}, \beta^{\text{ref}}(r)$ and  knots $\boldsymbol{\xi}^{\text{EA}} \}$.}
\Ensure{Optimum $\hat{\alpha},\hat{\beta}$ of $p(q|\alpha,\beta)$}
\For{$i=1, \ldots, N_{\text{sample}}$}
\State Evaluate a random data set: $\theta_i = \mathcal{N} (\bar{\theta}, 0.2 \; \bar{\theta}) $
\Procedure{Eval $q_i$}{$\theta_i$}
\State $\hat{a}_i \gets $ $QP(q)$ solution of Eq. \eqref{DLI_III_QP}
\State  $q_i \gets \min S(q)$
\EndProcedure
\EndFor
\State Fit (MCMC) the hyperprior $p(q|\alpha,\beta)$ to the $\{ q_i \} $ data set to obtain maximum likelihood values of $\hat{\alpha},\hat{\beta}$
\end{algorithmic}
\caption{Optimum smoothing}
\label{DLI_III_Opt_smooth}
\end{algorithm}

\begin{figure}
\centering
%height=0.3\textwidth, width=0.5\textwidth
%\showthe\columnwidth 
\includegraphics[width=\columnwidth]{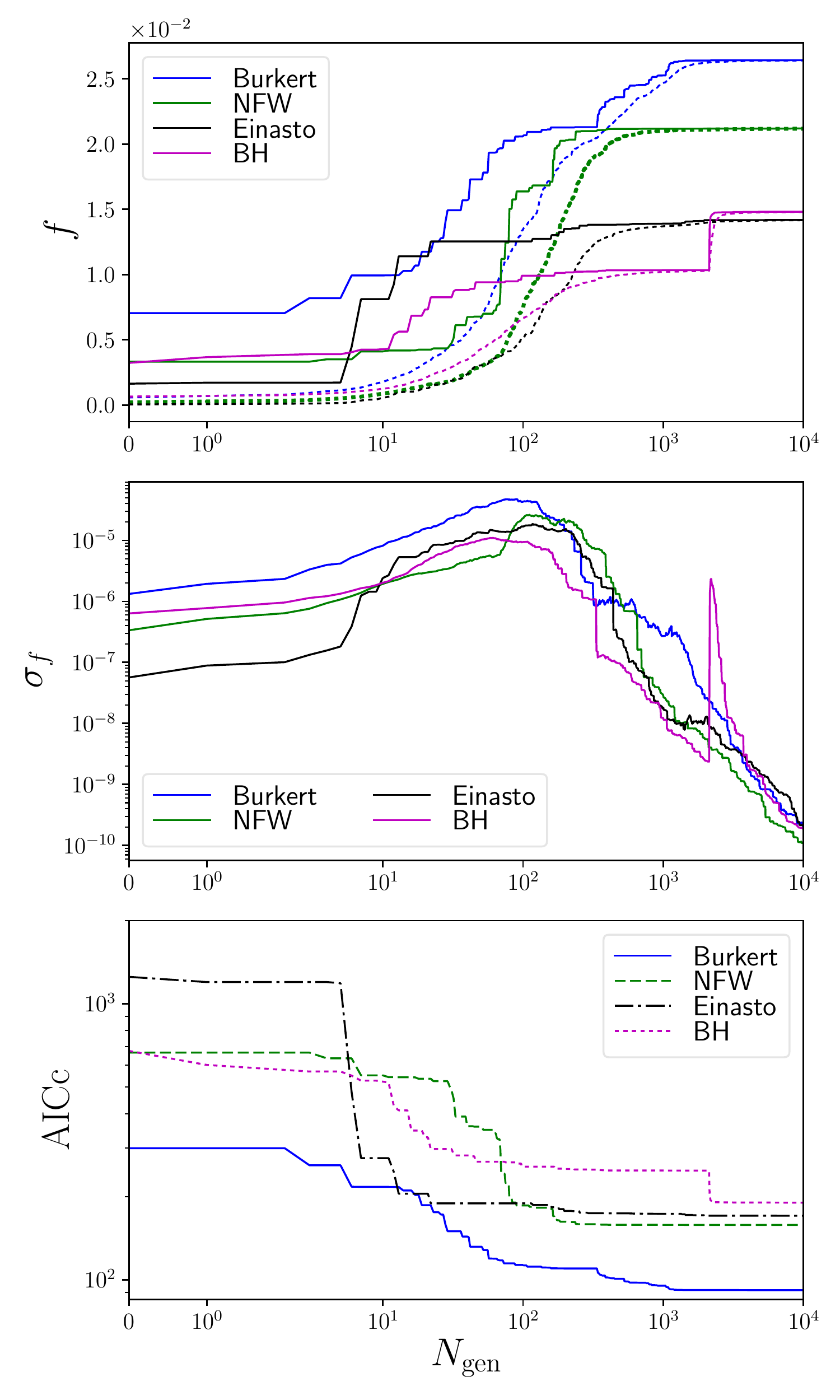}
\caption{EA optimization for various DM competing models for the synthetic data. Top panel: fitness value (Eq.   \eqref{DLI_III_fitFunction_general} and \eqref{DLI_III_AICc_general}) of the best individual (solid line) and average value of fitness (dotted line). Middle panel: standard deviation of the fitness values of the population. Bottom panel: AICc criterion of the  competing models.}
\label{DLI_III_FigEx1_comp}
\end{figure}

\subsection{Evaluation of marginalized distributions for the model parameters}
\label{DLI_III_sect_stat_anal}

Once we have selected an optimum knot distribution, we use Bayesian inference  in order to evaluate uncertainties of the various model parameters. 
For the evaluation of the marginalized distributions for our model parameters, we used an MCMC approach for the full likelihood in Eq. \eqref{DLI_III_loglkhood}. Specifically we designed a stepper that is a combination of Differential Evolution MCMC \citep{Braak:2006:MCM:1145406.1145416} and affine invariance MCMC \citep{GoodmanWeare} in a parallel tempering scheme. The library we used is \textsc{popmcmc++} (Diakogiannis 2016), a C++ library for population MCMC. It can be downloaded from https://github.com/feevos/popmcmc.

Since it is difficult to add hard bound constraints in the distribution function when performing MCMC sampling, the user of our method should carefully process the chains, after sampling, by removing possible model parameters that violate physical constrains (e.g. $\sigma_\text{rr}^2 < 0$).

\section{Example}
\label{DLI_III_Example_1}
In this section we will present an example of the usage of our algorithm. Initially the EA will evaluate the optimum knot distribution, $\boldsymbol{\xi}$,  as well as the best candidate DM model. For a robust modelling  we should also try various stellar models; however, for simplicity we restrict our analysis into varying only the DM models. Once we have the optimum EA knot distribution, $\boldsymbol{\xi}^{\text{EA}}$, and the best candidate model of DM, we use
MCMC to derive uncertainties in the model parameters. For simplicity, in this particular example we set $\Upsilon_V = 1$, i.e. we fit for the knot distribution, $\boldsymbol{\xi}$, the stellar, $\theta_{\star}$ and DM, $\theta_{\bullet}$, parameters. 

\subsection{Synthetic data}
We constructed a set of  synthetic data points  from a  \citet{1966AJ.....71...64K} brightness profile, a \citet{1995ApJ...447L..25B} DM mass density, and an MLT anisotropy profile 
\cite[][see also \citealt{2013MNRAS.429.3079M}]{2007A&A...476L...1T} defined by: 
\begin{equation}
\beta_{\text{MLT}}=\beta_{\infty} \frac{r}{r+r_{\beta}}
\end{equation}
Our choice of parameters that produces a relatively difficult profile that is not approximated well with a uniform knot distribution is: 
\begin{align*}
\theta_{\star}^{\text{ref}} &= \{w_0^{\text{ref}}=7.5,\rho_{0\star}^{\text{ref}}= 50\, (\text{M}_{\odot} \, \text{pc}^{-3}), r_{\mathrm{c}}^{\text{ref}}=100 \, (\text{pc}) \}\\ 
\theta_{\bullet}^{\text{ref}} &= \{\rho^{\text{ref}}_{0\bullet}=10\; (\text{M}_{\odot} \, \text{pc}^{-3}), r_{\mathrm{s}}^{\text{ref}}= 200\; (\mathrm{pc})  \} \\ \theta_{\beta} &= \{ \beta_{\infty} = -0.5,  r_\beta = 15 \, (\text{pc}) \}
\end{align*}
From these reference values, and in the range $x\in [0,r_{\text{t}}]$ 
( $r_{\mathrm{t}} \approx 1 \; (\mathrm{kpc})$) we chose 20 position data points, $x_i$, with a mild exponential concentration close to the origin. For each $x_i$ we created a random brightness, $\Sigma_i$, and a random line-of-sight velocity dispersion value, $\sigma^2_i$. 
These values are created with the same profiles but with random values on their defining parameters, $\theta^{\mathrm{rand}}_{\star}$ and $\theta_{\bullet}^{rand}$, for each $x_i$. These random parameters are Gaussian  random numbers centered on the reference values, $\theta^{\text{ref}}$, with variance equal to  $10\%$ of the reference value, $\sigma_{\theta} = 0.1 \theta^{\text{ref}}$, i.e. $\theta^{\mathrm{rand}} \sim \mathcal{N}(\theta^{\text{ref}}, \sigma_{\theta})$. The synthetic data set can be seen in Fig. \ref{DLI_III_FigEx1_sigmaLOS_1}. The difference in the magnitude of errors is a result of the normality of the errors of the model parameters.

\subsection{Results} 

\subsubsection{EA model selection}

In Fig \ref{DLI_III_FigEx1_comp} we plot the evolution of the fitness values for various competing DM models. 
These are Burkert, Einasto, NFW and a fit with a simple black hole in the center.   The horizontal axis represents the generations of the EA in log space. In the top panel we plot the fitness value, $f$, as a function of the generations, $N_{\text{gen}}$. Solid lines correspond to the maximum fitness value of the population, $f_{\text{max}}$, while dotted lines  to the average  fitness, $\langle f \rangle$, of the generation. As the generations evolve, the average fitness value approaches the fitness  
of the best individual. This is an indication of the convergence of the algorithm around the best solution. In the middle panel we plot the standard deviation of the fitness values of the population. The standard deviation, $\sigma_f$, initially rises, i.e. the spread of the fitness values increases. This is because the EA  tries to encapsulate in its population the best solution, therefore it spreads the values of the solutions in order to explore parameter space. As the EA progresses and the best candidate has been found and  is included in the population, the population tends to ``shrink'' around the best solution. This is why the average value gets closer to the highest fitness value, and the standard deviation gets minimized. Some spikes that are observed in the $\sigma_f$ are because of recombination: a better solution was found away from the median of the population and the EA spreads until it encapsulates it. For example, the sudden change in fitness and AICc of the BH model at generation $\approx 2000$ is accompanied by a spike in the dispersion of the fitness. 

In the bottom panel we plot the corresponding values of the AICc information criterion. 
Note that this AICc is evaluated from Eq. \eqref{DLI_III_AICc_general} with $n$ being the total number of parameters, i.e. $n=\dim(\theta_{\star},\theta_{DM},\{a_i \})$. This is why 
the highest fitness value does not necessarily correspond to the minimum AICc.
In Table \ref{DLI_III_synthetic_AICc} we report the AICc values, as well as the differences, $\Delta$, between the best AICc model (Burkert) with its competitors. For selecting the most probable model,  we follow \cite{opac-b1100695} who state that model selection between two competing models is conclusive if the difference of their AICc values is greater than 10. 
The Burkert DM model (the reference from which the synthetic data were created) 
is clearly distinguished from competing solutions, so the algorithm is successful in rejecting inappropriate  models.

\begin{table}
\caption{Synthetic data, AICc comparison of models. A value of $\Delta > 10$ states model selection conclusive 
\citep{opac-b1100695}.}
\label{DLI_III_synthetic_AICc}
\begin{center}
\begin{tabular}{ | l | l | l |  }
\hline
Model & AICc & $\Delta$ \\ \hline
King, Burkert & $91.76$ & $0.0$ \\ 
King, NFW & $158.15$  & $66.38$ \\ 
King,  Einasto & $170.72$ & $78.96$ \\ 
King, BH & $190.48$  & $98.72$ \\ 
\hline
\end{tabular}
\end{center}
\end{table}

\begin{figure}
\centering
%height=0.3\textwidth, width=0.5\textwidth
%\showthe\columnwidth 
%\includegraphics[width=\columnwidth]{images/FigEx1_BIN_G3PCX_1.pdf}
\includegraphics[width=\columnwidth]{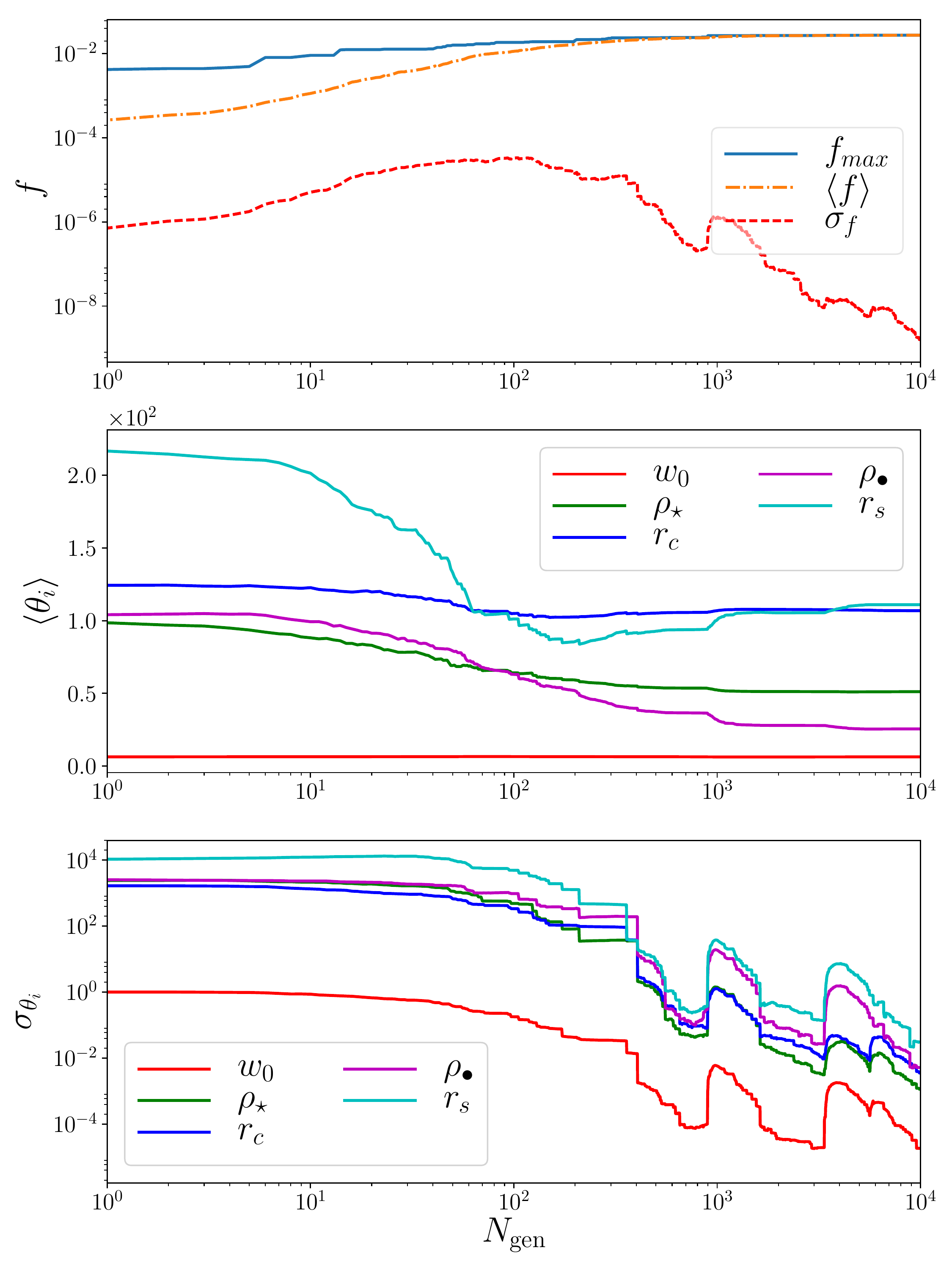}
\caption{Evolution of most favoured model for the synthetic data set. Top panel: maximum fitness value (solid line - Eq.   \eqref{DLI_III_fitFunction_general} and \eqref{DLI_III_AICc_general}), average fitness (dash-dotted line) and fitness variance (dashed line). Middle panel: evolution of the average value, $\langle \theta_i \rangle$, of the model parameters. Bottom panel: evolution of the variance, $\sigma_{\theta_i}$, of each of the model parameters. }
\label{DLI_III_FigEx1_1}
\end{figure}

In Fig \ref{DLI_III_FigEx1_1} we focus on the evolution of the EA of the Burkert DM model. In the top panel we plot the fitness of the best individual, $f_{\text{max}}$, the average fitness, $\langle f \rangle$, and the standard deviation, $\sigma_f$, of the fitness values. The vertical axis of the top figure is in log space. The horizontal axes are in log space for all panels.  In the middle panel we plot the evolution of the average value of each of the model parameters, $\langle \theta_i\rangle$, within each generation. 
The vertical axis of this panel is in linear space. Finally
 in the bottom panel we plot the standard deviation, $\sigma_{\theta_i}$, of the values of each parameter, $\theta_i$, in the population for each generation. As expected the  $\sigma_{\theta_i}$ tends to get smaller as the optimization evolves. Occasional spikes are produced as a result of the recombination operation: a better candidate is produced and the EA performs a small exploration of the solution space, before converging again.

\begin{figure}
\centering
%height=0.3\textwidth, width=0.5\textwidth
%\showthe\columnwidth 
%\includegraphics[width=\columnwidth]{images/EAfit_Ex1_sigmaLOS.pdf}
\includegraphics[width=\columnwidth]{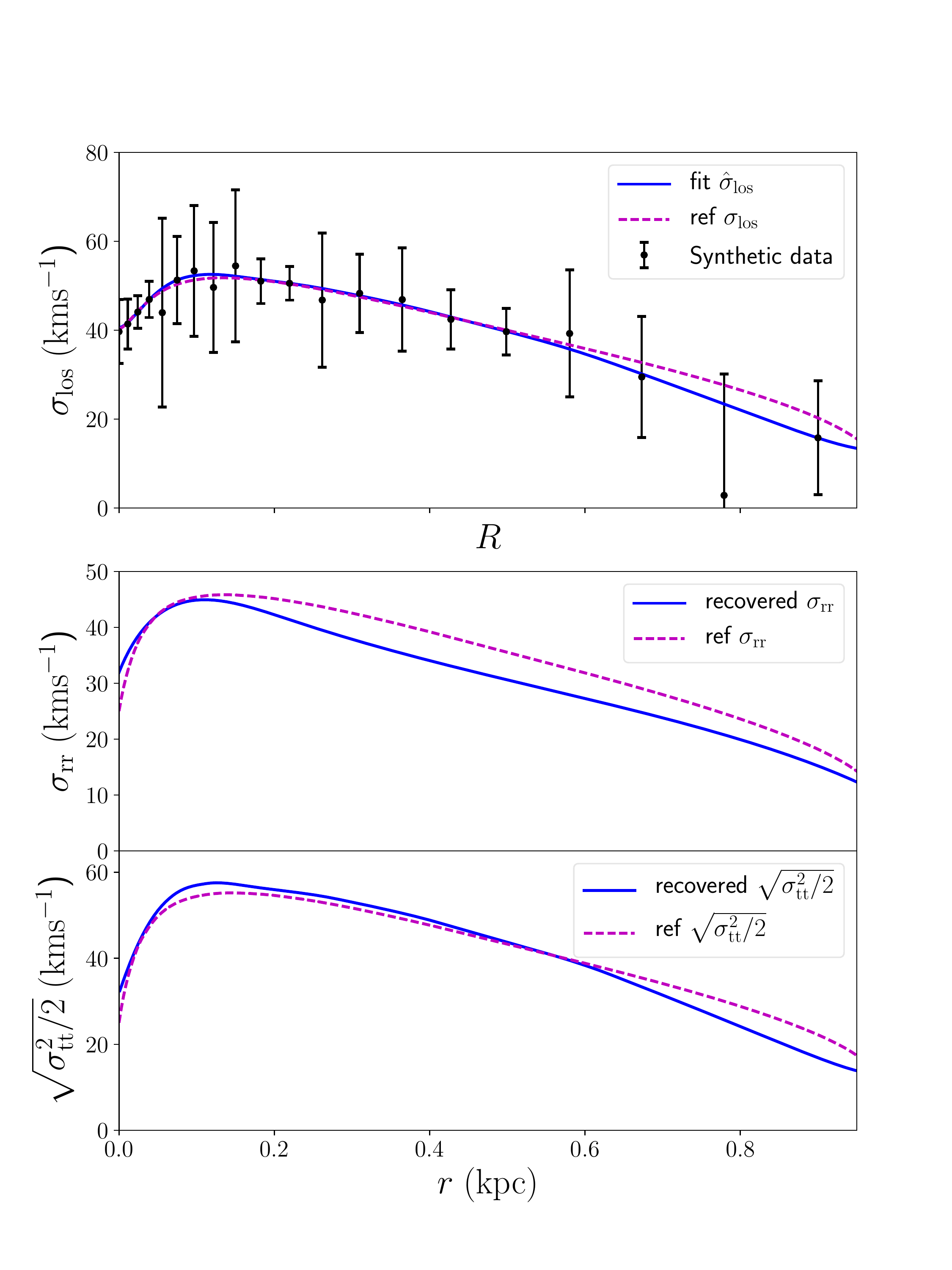}
\caption{EA best fit for the synthetic data set. Top panel: line-of-sight velocity dispersion, $\sigma_{\text{los}}$.  Middle panel: radial velocity dispersion, bottom: tangential velocity dispersion. In all panels the dashed lines correspond to the true values.}
\label{DLI_III_FigEx1_sigmaLOS_1}
\end{figure}

In Fig \ref{DLI_III_FigEx1_sigmaLOS_1} we plot the highest fitness solution as  evaluated from the EA. 
In the top panel we plot the line-of-sight velocity dispersion, $\sigma_{\text{los}}$ fit. In the same plot we also visualize the synthetic data set and the reference profile from which we created the synthetic data. 
In the middle panel we plot the  radial velocity dispersion as well as the reference profile. In the bottom panel we plot the tangential velocity dispersion. 
The EA is using the AICc as a means of optimum smoothing (instead of prior information from ideal theoretical models). This means that we may get fits that demonstrate non-physical oscillatory behaviour,  however the EA can identify the correct mass model as well as the region of interest for the best knot distribution. Optimum smoothing based on ideal theoretical models results in reliable estimates of the kinematic profiles after we have selected a good  knot distribution.

The EA recovered parameters are $\xi^{\text{EA}}=\{0,0.21,1 \}$ and  $\{ w_0^{\text{EA}}=6.19,\rho_{0\star}^{\text{EA}}=  51.46 \, (\text{M}_{\odot} \; \text{pc}^{-3}), r_{\mathrm{c}}^{\text{EA}}=  107.13\; (\mathrm{pc})$, 
$\rho_{0\bullet}^{\text{EA}}=  27.72 \; (\text{M}_{\odot} \, \text{pc}^{-3}),  r_{\mathrm{s}}^{\text{EA}}=  105.04\; (\mathrm{pc}),  q^{\text{EA}}=  0.52 \}$.
  The knots $\xi_i^{\text{EA}}$ are evaluated in the unit interval and must be scaled to $[0,r_{\text{t}}]$ before we can use them. Our estimated parameter values are close except for the $w_0^{\text{EA}}$ and the DM model parameters.  We cannot make any prediction at this stage of the deviation from the true values since we have no uncertainty for our estimates. After all, we are only interested in selecting the best candidate mass model, as well as identifying a really good knot distribution that will allow us to perform a robust statistical fit.

\begin{figure}
\centering
%height=0.3\textwidth, width=0.5\textwidth
%\showthe\columnwidth 
%\includegraphics[width=\columnwidth]{images/FigEx1_IGab_params.pdf}
\includegraphics[width=\columnwidth]{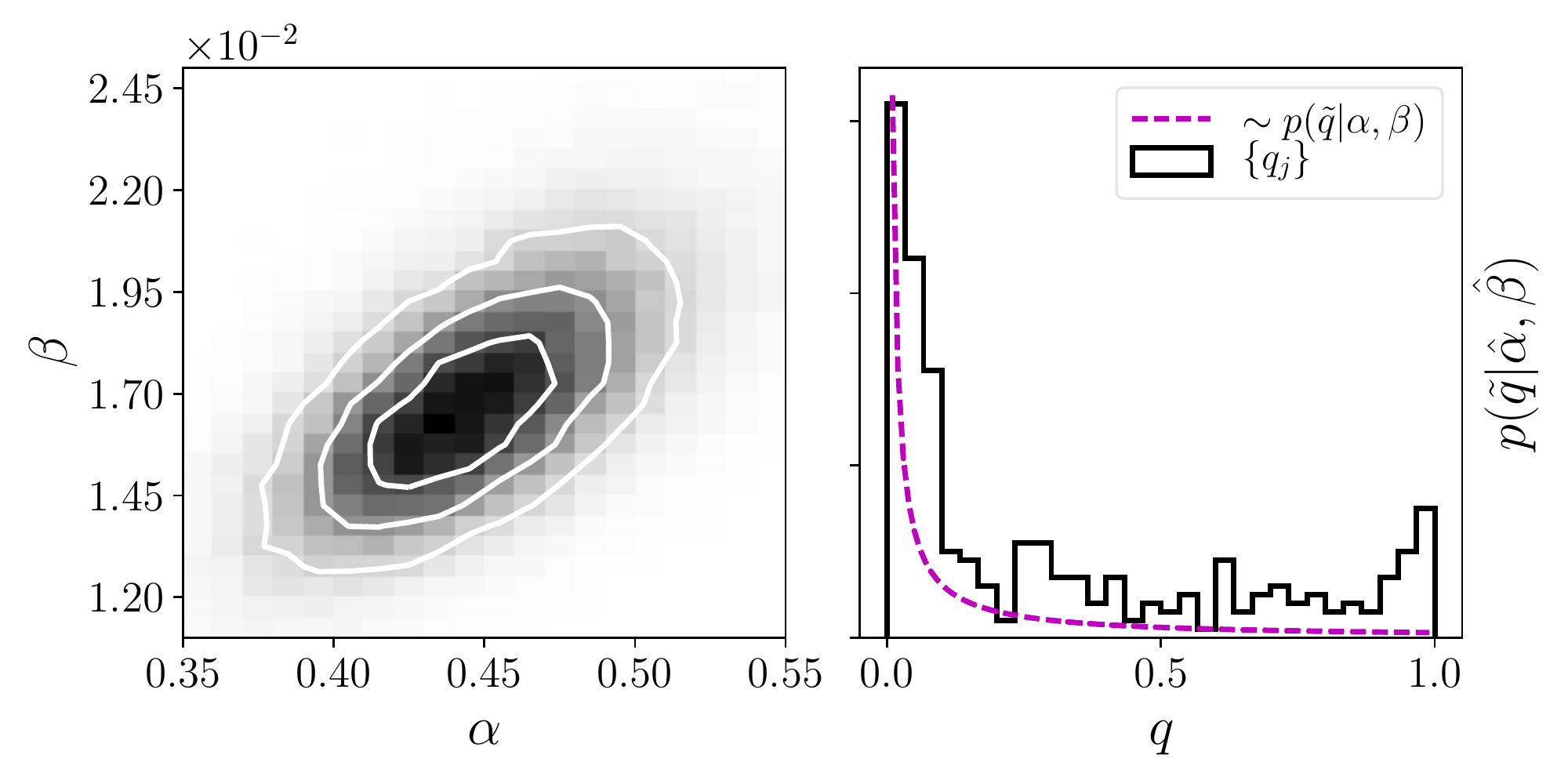}
\caption{Left panel: Marginalized distributions of $\alpha,\beta$ parameters of hyperprior $p(\tilde{q}(q)|\alpha,\beta)$ for the synthetic data set. Right panel: histogram of  data values $\{ q_j\}$ as these were evaluated from  mock data sets of ideal theoretical models. Overplotted is the  the hyperprior 
$p(\tilde{q}(q)|\hat{\alpha},\hat{\beta})$ distribution for the maximum likelihood estimated parameters $(\hat{\alpha},\hat{\beta})$. }
\label{DLI_III_FigEx1_2}
\end{figure}

\subsubsection{Confidence interval for model parameters}

Once we have chosen a suitable  knot vector, $\boldsymbol{\xi}$, we can focus on repeating the fit with MCMC, keeping the knots, $\xi_i$,   fixed  so as to derive uncertainties on the model parameters,  $\{\theta_{\star}, \theta_{\bullet}, a_i, q\}$. In order to do so we first need to estimate the parameters $\alpha, \beta$ of the hyperprior $p(\tilde{q}|\alpha,\beta)$ that regulate the smoothing. For this we apply Algorithm \ref{DLI_III_Opt_smooth} (see Sect. \ref{DLI_III_OptSmoothing_sec}).

\subsubsection{Evaluation of $\alpha,\beta$ smoothing parameters}

Our goal is to estimate the $\alpha,\beta$ parameters of the hyperprior $p(\tilde{q}|\alpha,\beta)$. In order to do so we need to obtain a  meaningful (for the problem in hand) set of $N$ values $q_j$ and then find the values  $\hat{\alpha},\hat{\beta}$ that maximize the likelihood:
\begin{equation}
\mathcal{L}_q = \prod_{j=1}^{N}  p(\tilde{q}(q_j)|\alpha,\beta). 
\end{equation}
Any optimization algorithm will do, we choose MCMC for this task. 
According to   Algorithm \ref{DLI_III_Opt_smooth} we need to create: a) An ideal reference  profile that will be used for the evaluation of $a_i^{\text{ref}}$ values, and b) a realization of $N$ synthetic data sets, $\theta^j = \{\theta_{\star}^j,  \theta_{\bullet}^j\}$ that each will give us a value $q_j$ that minimizes the measure $S(q)$ (Eq. \ref{DLI_III_smoothing_pnlty}). 
As an ideal reference model\footnote{This is used for the evaluation of $a_i^{\text{ref}}$, see Eq. \eqref{DLI_III_smoothing_pnlty}. } we use the EA solution based on the real data set as well as the  Osipkov-Merritt anisotropy profile, $\beta(r)$ (Eq. \ref{DLI_III_OM}), with $c_1=1$ and $c_0 = \xi_2\, r^{\text{EA}}_{\mathrm{t}}$, where $r^{\text{EA}}_{\mathrm{t}}$ is the  EA evaluated tidal radius, $r^{\text{EA}}_{\mathrm{t}}$ from the best fitting King profile.

In Fig. \ref{DLI_III_FigEx1_2} in the left panel we plot the Markov Chains of the $\alpha$, $\beta$ parameters as these were obtained from the MCMC algorithm. In the right panel we plot the maximum likelihood (scaled) hyperprior  $p(\tilde{q}|\hat{\alpha},\hat{\beta})$ as well as the histogram that corresponds to 300 estimated values $q_j$ that minimize the function $S(q)$ (Eq. \ref{DLI_III_smoothing_pnlty}).

Once we have the maximum likelihood values, $\hat{\alpha},\hat{\beta}$, we substitute them in  Eq. \eqref{DLI_III_loglkhood} and with an MCMC we obtain the Markov Chains of the model parameters. We use the restriction that $\sigma_{\text{rr}}(r_{\text{t}}) \approx \sigma_{\text{tt}}(r_{\text{t}}) \approx \sigma_{\text{los}}(r_{\text{t}}) \approx 0$. From an empirical point of view, close to the tidal radius, for bound stars, the radial velocity must approach zero ($r_{\text{t}}$ is a turning point for all bound particles approaching it). Hence we expect that the uncertainty in $v_r$ values is very small, this should lead to $\sigma_{\text{rr}}(r_{\text{t}})\to 0$. On the other hand, if we assume that there are purely tangential motions,  these must have very small velocity in order to remain in bound orbits (due to the small value of the gravitational force at $r=r_{\text{t}}$), i.e. $v_t \approx 0$. Hence we expect again that $\sigma_{\text{tt}}(r_{\text{t}})\to 0$. See also \cite{2013degn.book.....M} for a theoretical proof that as $r\to \infty$ this restriction is equivalent to the virial theorem for a self gravitating system in equilibrium. For the case where we wish to use a stellar model different than the King profile, the tidal radius can be evaluated from the virial radius of the system, i.e. $r_{\text{t}} \approx r_{\text{vir}}$. Our restriction on  the various velocity dispersions at $r_{\text{t}} = r_{\text{vir}}$ remains a good approximation.

\begin{figure*}
\centering
%height=0.3\textwidth, width=0.5\textwidth
%\showthe\columnwidth 
%\includegraphics[width=\textwidth]{images/Ex1_mcmc_chains.pdf}
\includegraphics[width=\textwidth]{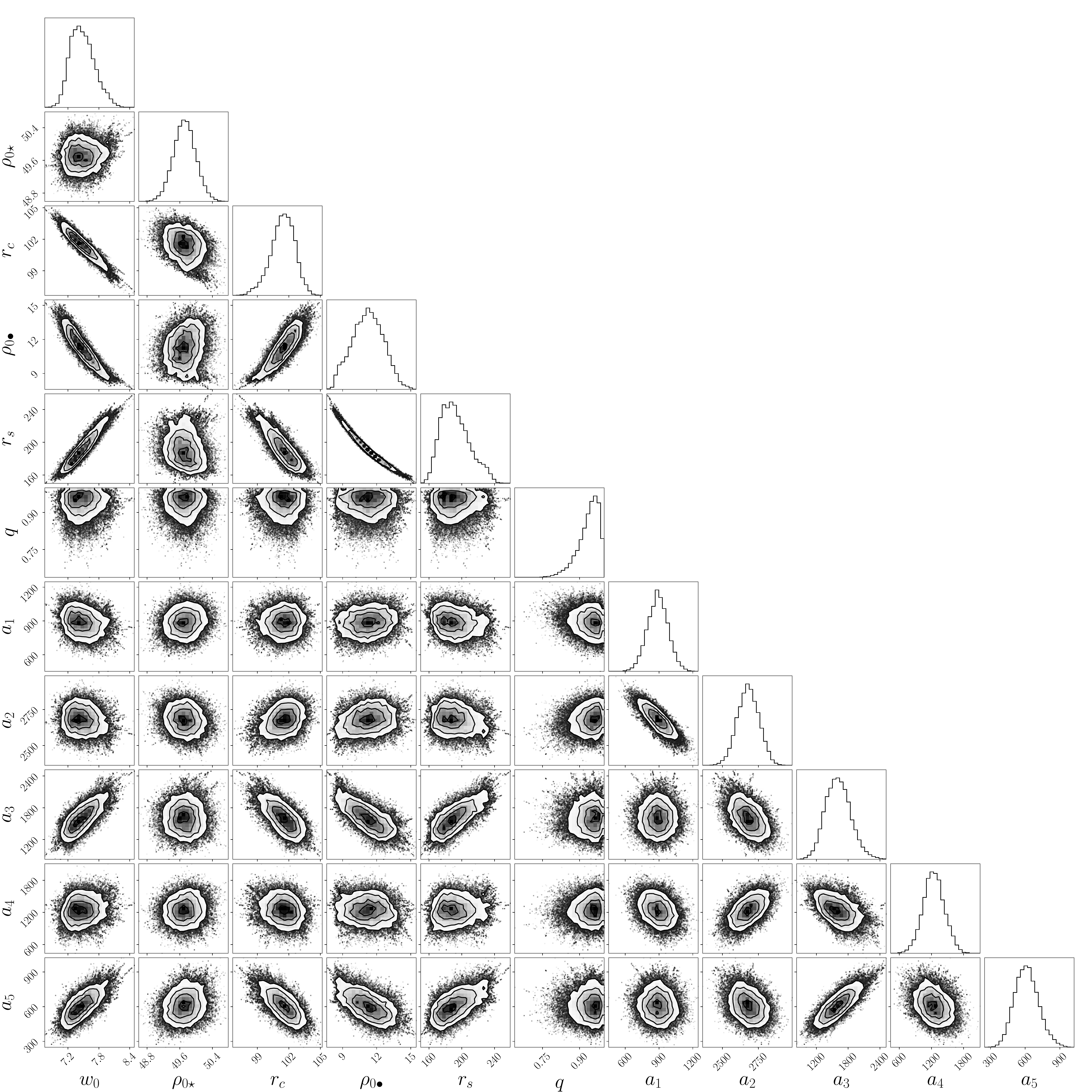}
\caption{MCMC marginalized distributions of the the Burkert  model parameters for the example with synthetic data.}
\label{DLI_III_mcmc_chains}
\end{figure*}

In Fig \ref{DLI_III_mcmc_chains} we plot\footnote{We used the python library \textsc{corner} from \cite{dan_foreman_mackey_2016_45906}.} the Markov Chains of the model parameters as obtained from the MCMC. In all marginalized distributions of the Markov chains the reference values of the profile that created the synthetic dataset are captured. 
There is a strong correlation between the $\rho_{0\bullet}$ and $r_{\mathrm{s}}$ parameters and this is one reason why the evolutionary algorithm gives the greatest error in the estimation of these values. This may have  also affected the $w_0^{\text{EA}}$ evaluated parameter that lies outside the marginalized distribution of the MCMC chains. Since the EA uses AICc  for the best model fit, it cannot account for the optimum smoothing from mock stellar systems that MCMC has built in its prior information. This is a source of bias in  the EA estimates,  but in all of our tests with mock data sets it does not seem to affect the model selection.

\begin{figure}
\centering
%height=0.3\textwidth, width=0.5\textwidth
%\showthe\columnwidth 
%\includegraphics[width=\columnwidth]{images/mcmcfit_Ex1_sigmaLOS.pdf}
\includegraphics[width=\columnwidth]{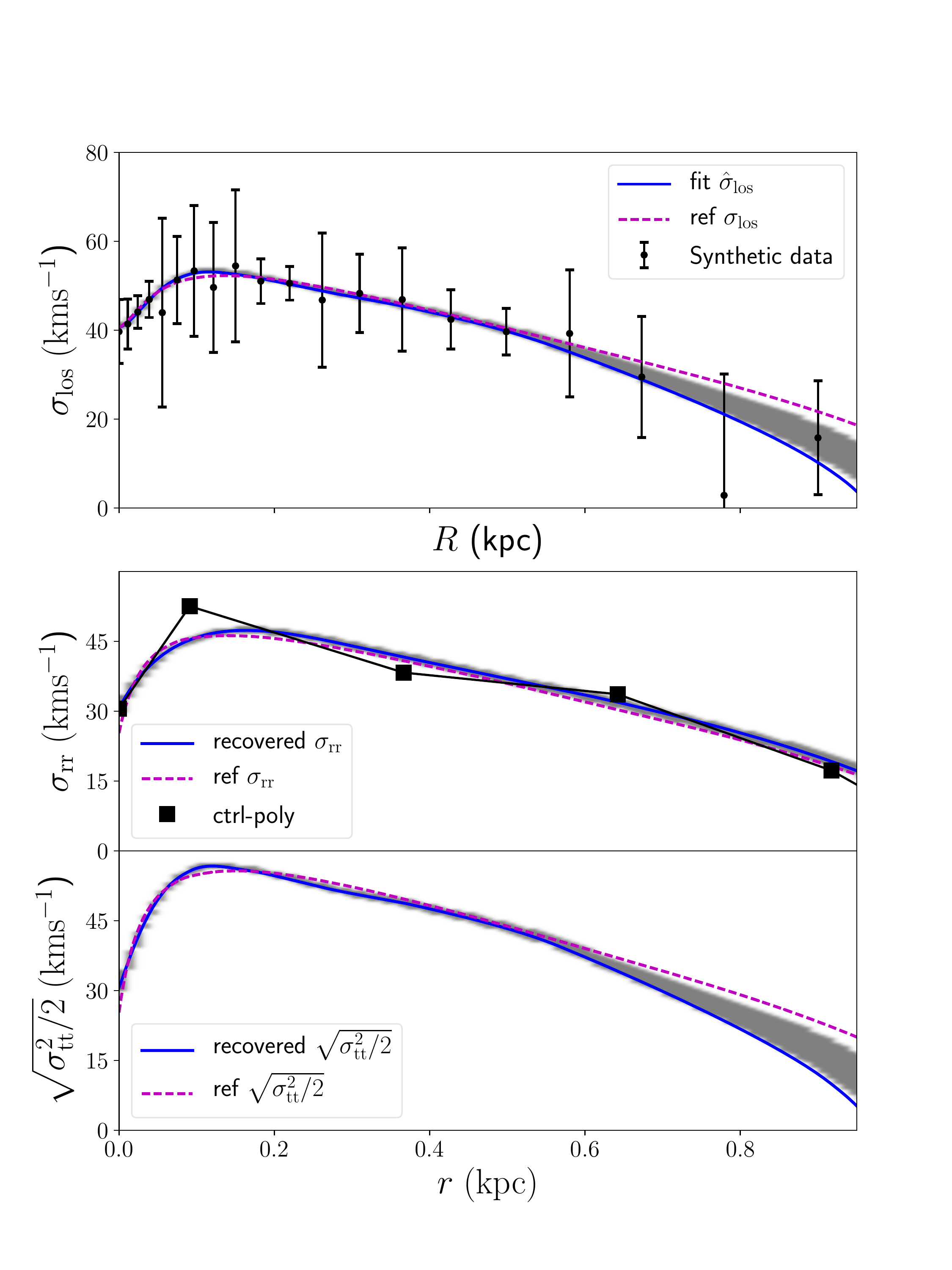}
\caption{MCMC highest likelihood fit (blue curves) of second order velocity moments for the synthetic data set. The grey region corresponds to 1$\sigma$ uncertainty region in all model parameters. The dashed lines corresponds to the true values of the reference profile from which the synthetic data were created. Black squares are the control points of the control polygon (solid black line) - see \citet{2014MNRAS.443..598D} for definition and details. }
\label{DLI_III_FigEx1_mcmc_fit}
\end{figure}

In Fig \ref{DLI_III_FigEx1_mcmc_fit} we plot the fit of the various velocity moments. In the top panel we see the line-of-sight velocity dispersion, $\sigma_{\text{los}}$, the true reference profile that corresponds to the synthetic data set, as well as the synthetic data. In the middle panel we plot the maximum likelihood radial velocity dispersion, $\sigma_{\text{rr}}$, and the corresponding reference value. We also show the control polygon that regulates the geometric shape of the B-spline $\sigma_{\text{rr}}$ function. Observe that there are two control points with $r$-coordinate closer to the origin, $r=0$, something evaluated by the EA. 
In the bottom panel we plot the tangential velocity dispersion, 
$\sqrt{\sigma_{\text{tt}}^2/2}=\sqrt{(\sigma_{\theta\theta}^2+\sigma_{\phi\phi}^2)/2}$. In all panels the grey region corresponds to the  1$\sigma$ uncertainty interval of the estimated values of all model parameters. The reconstruction is satisfactory. Even in the case where the fitted profile deviates from the true one (top and bottom panels for $R,r>0.6 \; \mathrm{kpc}$), the uncertainty is greater towards the true reference profile. Note that there is a difference in the best solution found by the EA (Fig. \ref{DLI_III_FigEx1_sigmaLOS_1}) and the MCMC (Fig. \ref{DLI_III_FigEx1_mcmc_fit}). The reason is that in the likelihood used in the MCMC we have encoded the smoothing information from ideal mock systems. This information is not available for the EA runs.

\begin{figure}
\centering
%height=0.3\textwidth, width=0.5\textwidth
%\showthe\columnwidth 
%\includegraphics[width=\columnwidth]{images/mcmc_anisotropy_fit_Ex1_sigmaLOS.pdf}
\includegraphics[width=\columnwidth]{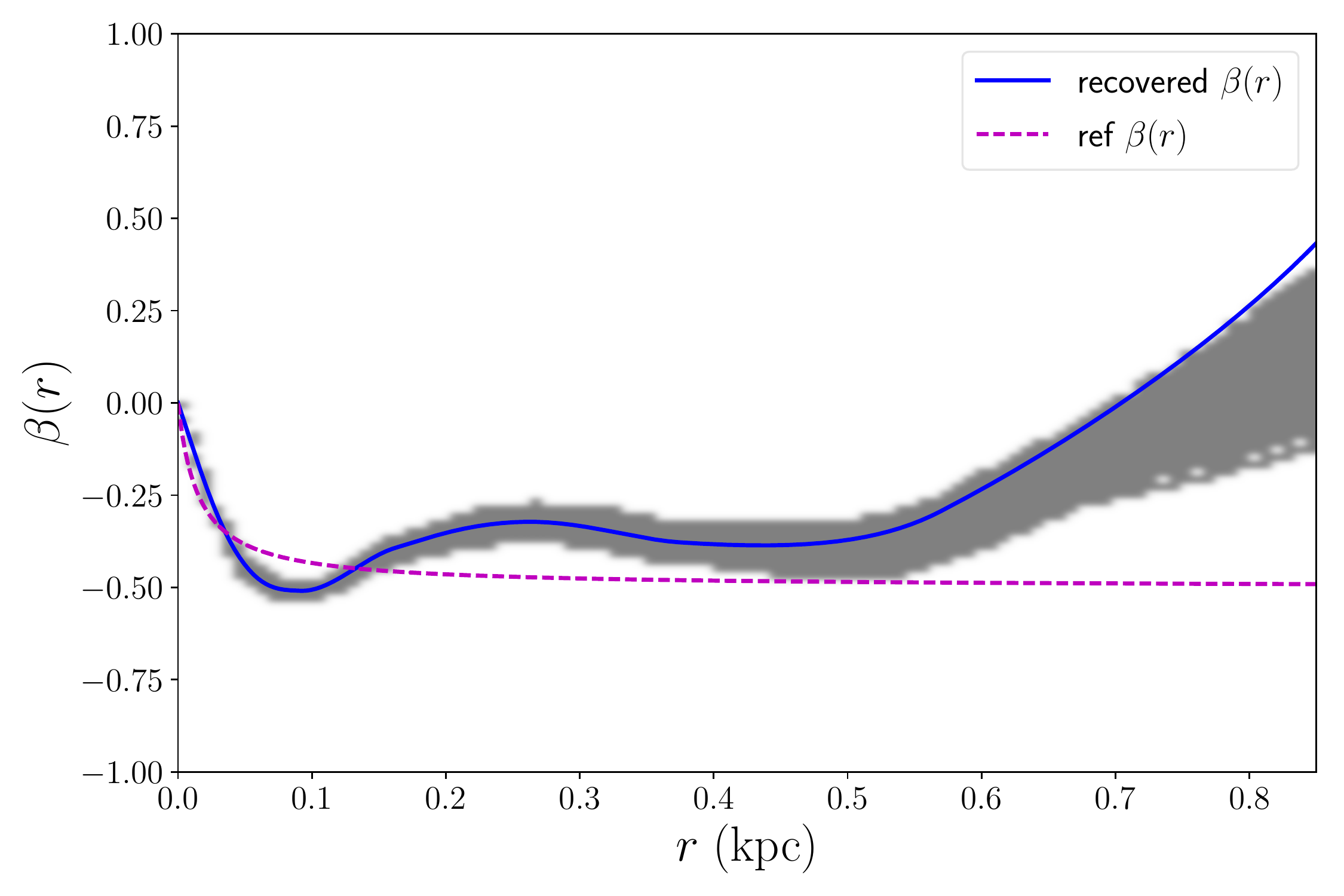}
\caption{ As Fig. \ref{DLI_III_FigEx1_mcmc_fit}, but for the anisotropy profile  $\beta(r)=1-\sigma_{\text{tt}}^2/(2\sigma_{\text{rr}}^2)$ for the synthetic data set.}
\label{DLI_III_FigEx1_2_anisotropy}
\end{figure}

In Fig.  \ref{DLI_III_FigEx1_2_anisotropy}  we plot the estimated anisotropy profile, $\beta(r)$ as well as the true reference anisotropy, $\beta_{\text{MLT}}(r)$ (recall $r_{\mathrm{t}} \approx 1 \; \mathrm{kpc}$).   The fact that we do not fit directly a functional form for $\beta(r)$ has several benefits and a penalty: the uncertainty in the $\beta(r)$ estimates is large, especially  for $r> 0.6 \; \mathrm {kpc} $. This is a result of the functional form that defines $\beta(r)$: the division  of $\sigma_{\text{tt}}^2$ by $\sigma_{\text{rr}}^2$ propagates large errors, especially closer to  the tidal radius, where both  $\sigma_{\text{rr}}^2$ and $\sigma_{\text{tt}}^2$ tend to zero.

\begin{figure*}
\centering
%height=0.3\textwidth, width=0.5\textwidth
%\showthe\columnwidth 
\includegraphics[width=\textwidth]{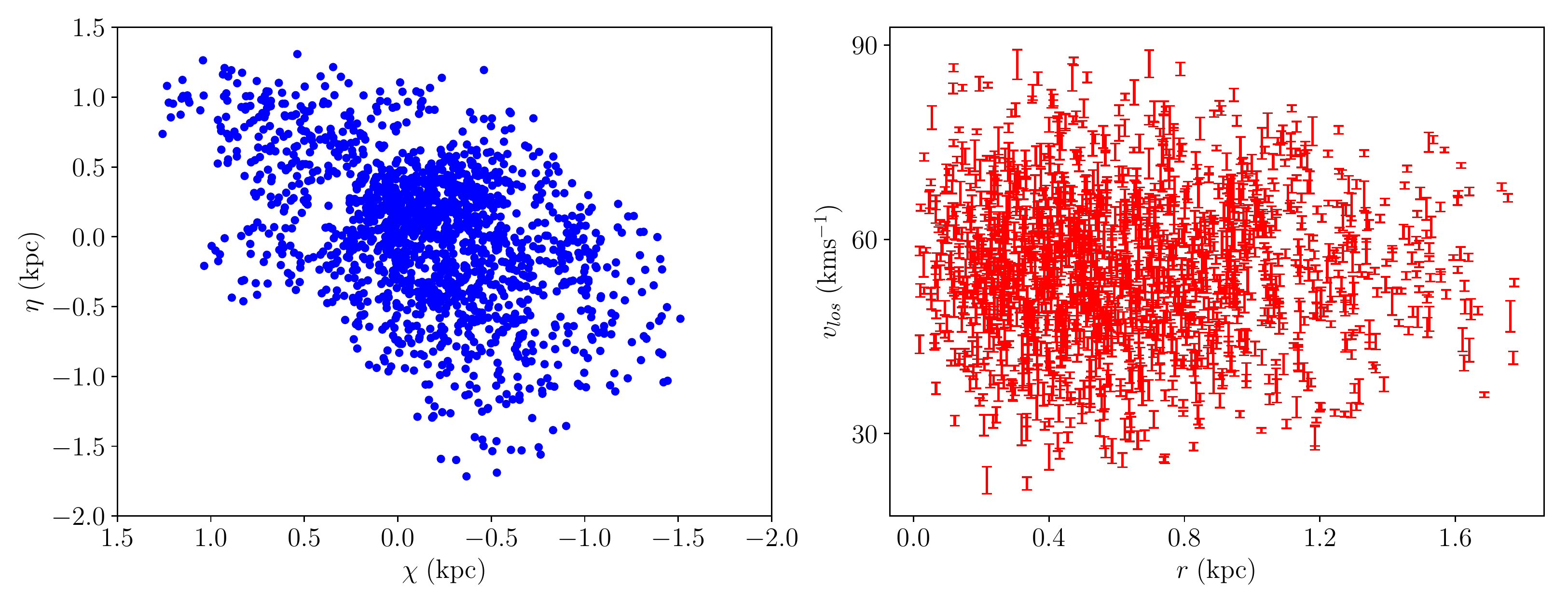}
\caption{Fornax member stars projected on the tangent plane (left) and their line-of-sight velocities (right). }
\label{DLI_III_fornax_tangent_plane_Vlos}
\end{figure*}

\section{Application to the Fornax dwarf spheroidal}
\label{DLI_III_Fornax_modelling}

In this section we apply the \textsc{JEAnS} algorithm to the Fornax dwarf spheroidal galaxy and present our  results. In Appendix \ref{DLI_III_Fornax_uniform_knots} we give some further insight on differences between EA adaptive knot modelling and modelling with  a uniform knot distribution.

\subsection{Data}
Fornax is a  dSph galaxy of the Local Group. It appears to be slightly flattened, with ellipticity $\epsilon \approx 0.3$ \citep{2015MNRAS.454.2401B}. It is at a distance $d=138 \pm 8 \; \mathrm{kpc}$ \citep{1998ARA&A..36..435M}  from the Sun and the coordinates of its centre are   RA: $2^{\text{h}} 40^{\text{m}} 04^{\text{s}}$, DEC: $-34^o 31' 30''$   \citep[J2000;][]{2003A&A...406..847W, 2005AJ....129.1443C}. We summarize this information in Table \ref{DLI_III_Fornax_params}.
We used published heliocentric velocity values and membership characterization from
 \cite{2009AJ....137.3100W, 2009AJ....137.3109W}. 
Targets were 
considered  members  if their  membership probability was greater than 0.8. This resulted in total $2248$ Fornax members.  
For the brightness, we constructed the  projected number density profile, normalized to the total luminosity of Fornax, $L_V = (1.79 \pm 0.69)\times 10^7 \, \text{L}_{\odot}$  
\citep[][based on \citealt{1995MNRAS.277.1354I} and \citealt{1998ARA&A..36..435M}]{2009MNRAS.394L.102L}. We did not account for the error in the $L_V$ value, instead the errors were estimated using the bootstrap method. 
In order to estimate the velocity dispersion values, we binned the data in 19 variable size circular annuli bins, each one containing 82 stars.  
We estimated the line-of-sight velocity dispersions
in each radial bin via our MCMC scheme (assuming again Gaussian line-of-sight velocity
distributions). 
In Fig. \ref{DLI_III_fornax_tangent_plane_Vlos} we plot the positions of the Fornax members on the tangent plane (left panel), and the heliocentric line-of-sight velocities with respect to the distance from the cluster centre (right panel).

\begin{table}
\caption{Fornax parameters \citep{2009MNRAS.394L.102L}}
\label{DLI_III_Fornax_params}
\begin{center}
\begin{tabular}{@{} | c | c | c | }
 \hline
 Centre (J2000) & $L_V (10^7 \, \text{L}_{\odot})$  & 
 Distance ($\mathrm{kpc})$ \\\hline
 RA: $2^{\mathrm{h}} 40^{\mathrm{m}} 04^{\mathrm{s}}$ & $1.79 \pm 0.69  $  & $d = 138 \pm 8 $   \\ 
 DEC: $-34^o 31' 30''$  & & \\ \hline
  \end{tabular}
\end{center}
\end{table}

\subsection{Results}

We model Fornax by assuming a \cite{1966AJ.....71...64K} brightness profile, a constant $V$-band mass-to-light ratio, $\Upsilon_V$, and in addition the following separate DM components: NFW, Burkert, Einasto and a black hole in the centre of the galaxy. In all of the models there is a constant mass-to-light ratio, $\Upsilon_V$, but only in one we do not use a separate DM mass profile, thus in total we have 5 different mass models. We refer to the model with no separate DM component as const$\Upsilon_V$.  In Table \ref{Fornax_range_params} we give the range of the model parameters that we used in the evolutionary algorithm. Clearly our range of parameters range from unrealistically small to unrealistically large mass models.
\begin{table}
\caption{Range of parameters for the various models we used inside the evolutionary algorithm for the case of Fornax.}
\begin{center}
\begin{tabular}{|l|l|l|}
\hline 
Model    & Parameter & $\theta$ range  \\ \hline
King & $W_0$ & $[0.5 ,20] $    \\
& $\rho_{\star 0}$ & $[10^{-5},0.4]$ $\text{M}_{\odot} \text{pc}^{-3}$\\
& $ r_{\mathrm{c}} $  & $[100,1500]$ pc \\ \hline
Burkert + NFW & $\rho_{\bullet 0}$ & $[0.0,0.4]$  $\text{M}_{\odot} \text{pc}^{-3}$ \\
& $r_{\mathrm{s}}$ & $[100.0,2000.0]$ pc\\ \hline
Einasto & $\rho_{\mathrm{e}}$ & $[0.0,2.0]$ $\text{M}_{\odot} \text{pc}^{-3}$\\
& $r_{\mathrm{e}}$ & $[2.0,2000.0]$ pc\\
 & $n$ & $[1,10]$ \\ \hline
Black Hole & $M_{\bullet}$ & $[10^2,10^6]$ $\text{M}_{\odot}$\\ \hline
Mass-to-light ratio & $\Upsilon_V$ & $[0.2,100]$ $\text{M}_{\odot} / \text{L}_{\odot}$  \\ \hline
 \end{tabular}
 \end{center}
\label{Fornax_range_params}
\end{table}
Despite the elliptical shape of Fornax,  we make the simplifying assumption that it is a spherically symmetric object in virial equilibrium that is not affected by the tidal field of the host. This seems a viable  scenario \citep{2009MNRAS.394L.102L,2015MNRAS.454.2401B}.   In our analysis we do not account  for contamination from binaries,  however numerical simulations demonstrate that this should not have a significant effect in our dispersion estimations \citep[][and references therein]{2012ApJ...746...89J}.

\begin{figure}
\centering
%height=0.3\textwidth, width=0.5\textwidth
%\showthe\columnwidth 
%\includegraphics[width=\columnwidth]{images/FigFornaxCompet_1.pdf}
\includegraphics[width=\columnwidth]{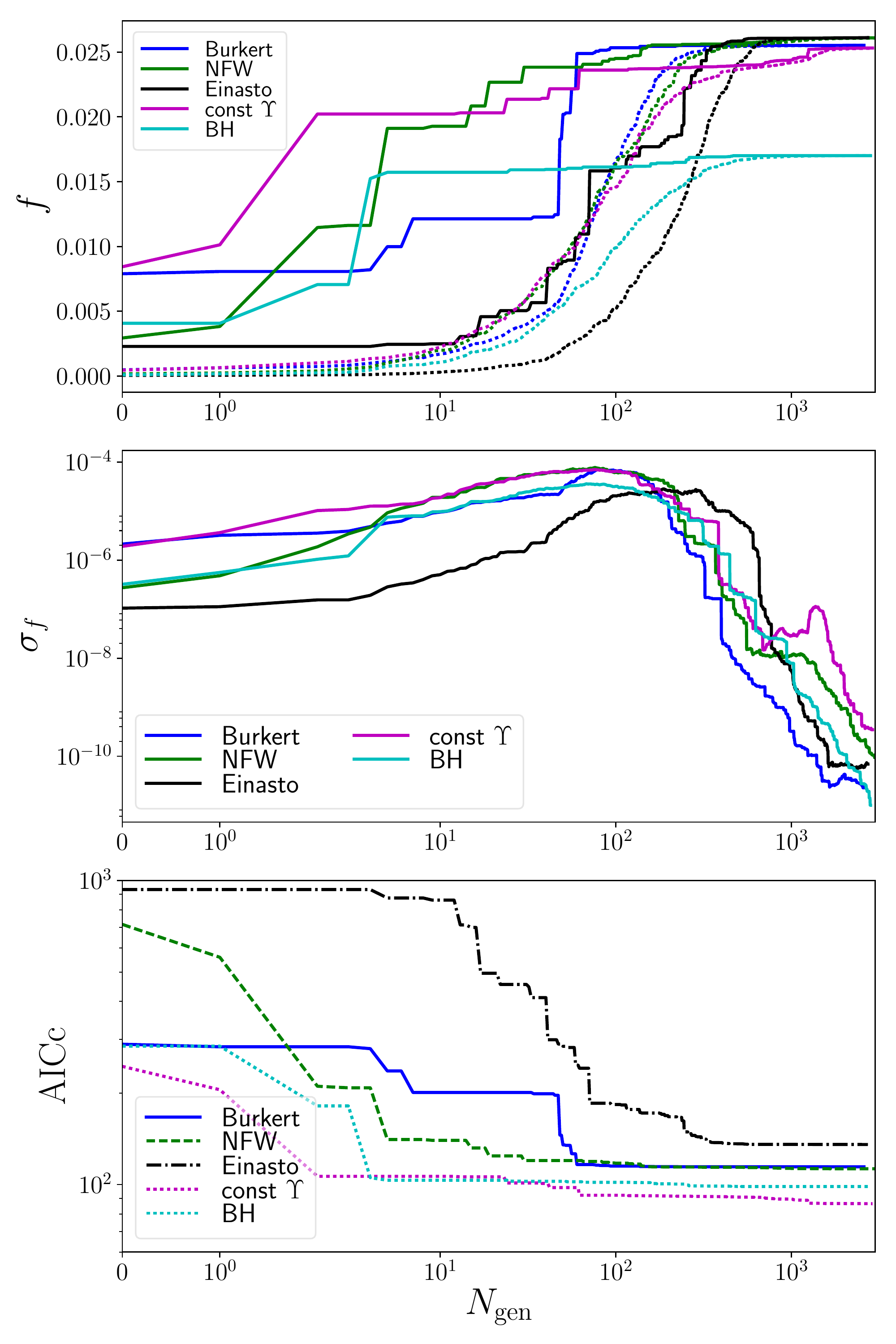}
\caption{
EA optimization for various DM competing models for the Fornax dSph. Top panel: fitness value (Eq.   \eqref{DLI_III_fitFunction_general} and \eqref{DLI_III_AICc_general}) of the best individual (solid line) and average value of fitness (dotted line). Middle panel: standard deviation of the fitness values of the population. Bottom panel: AICc criterion of the  competing models.
}
\label{DLI_III_FigFornax_1}
\end{figure}

\begin{figure}
\centering
%height=0.3\textwidth, width=0.5\textwidth
%\showthe\columnwidth 
%\includegraphics[width=\columnwidth]{images/FornaxCompet_2.pdf}
\includegraphics[width=\columnwidth]{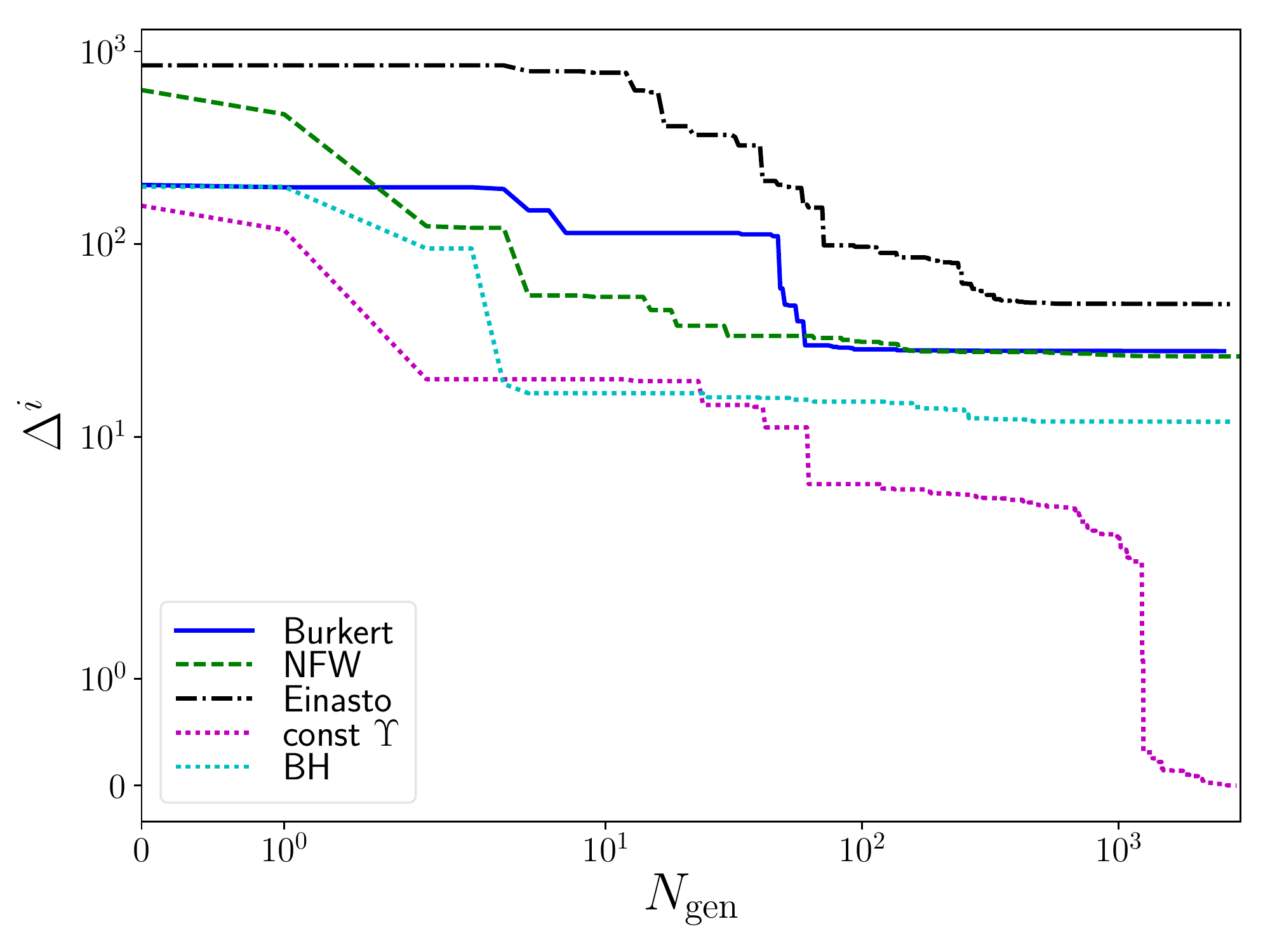}
\caption{Comparison of the AICc convergence criterion for the Fornax dSph. The differences $\Delta^i = \text{AICc}^i - \min (\{ \text{AICc} \})$ are plotted. The case of const$\Upsilon_V$ (simple King profile with constant mass-to-light ratio and no separate DM component) is clearly favoured over all competing models ($\Delta > 10$).}
\label{DLI_III_FigFornax_2}
\end{figure}

\begin{figure}
\centering
%height=0.3\textwidth, width=0.5\textwidth
%\showthe\columnwidth 
%\includegraphics[width=\columnwidth]{images/FornaxMass_n_light.pdf}
\includegraphics[width=\columnwidth]{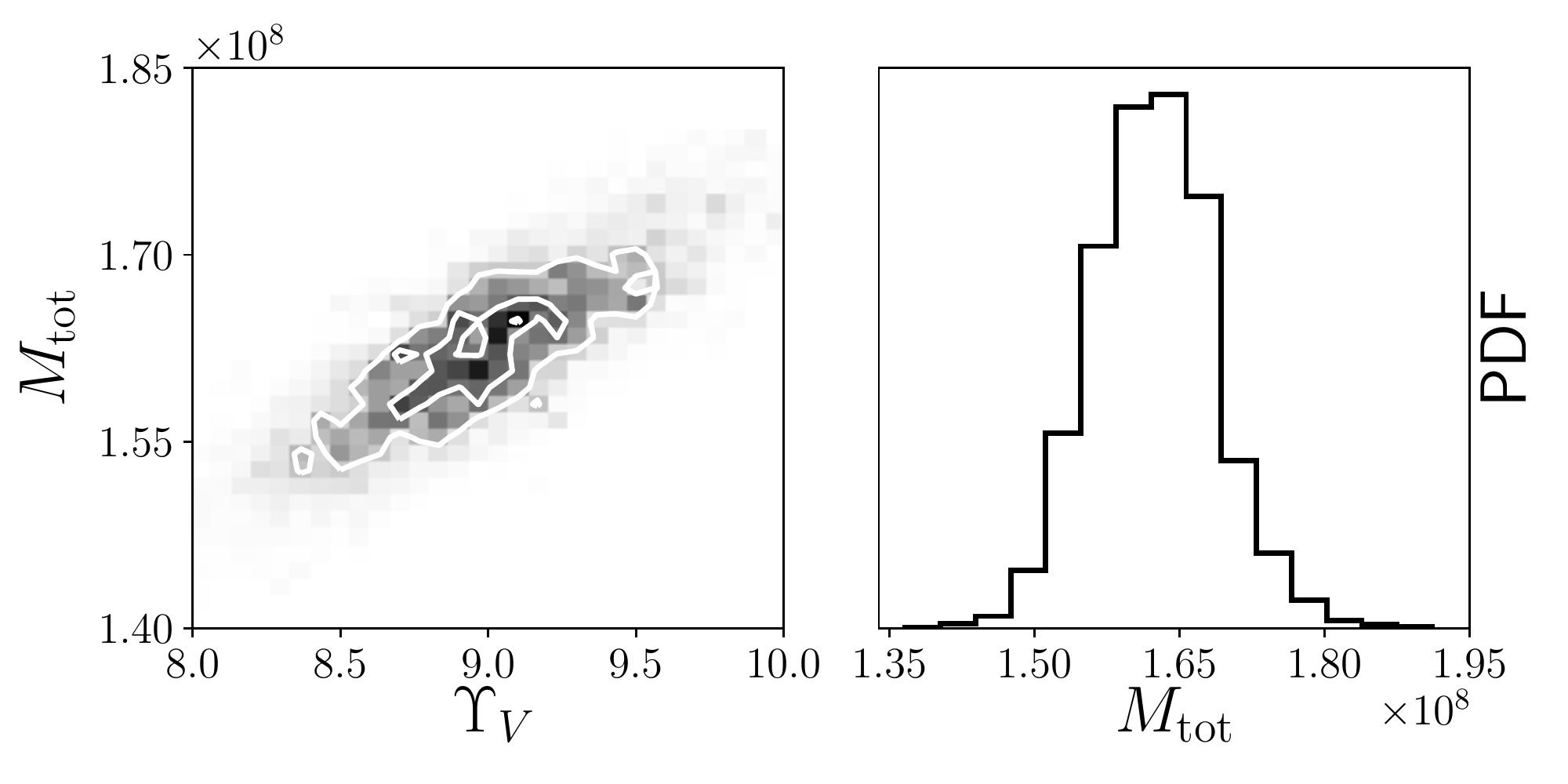}
\caption{Mass-to-light ratio, $\Upsilon_V$ and total mass, $M_{\text{tot}}$, as estimated from the MCMC chains for the Fornax dSph.}
\label{DLI_III_FigFornax_5}
\end{figure}

We start by running the EA for the five different models. 
The set up for the EA is the same as in the example with the mock dataset (G3 evolution model, $\mu = 5, \lambda = 20, p_{\mathrm{m}} = 0.005$, see Sect. \ref{DLI_III_EvoMod}). We evolved the algorithm for 3500 generations; this was adequate enough for convergence.  
In Fig. \ref{DLI_III_FigFornax_1} we plot (top panel) the evolution of the maximum fitness value (solid lines), as well as the average fitness for the population (dotted lines). As the algorithm converges around the global maximum the average value, $\langle f \rangle$, converges to the maximum fitness, $f_{\text{max}}$. In the middle panel we show the evolution of the dispersion in the fitness values, $\sigma_f$. After approximately 100 generations the algorithm has estimated a global maximum and from that point and on  the dispersions decrease gradually. Finally in the bottom panel we plot the AICc. We emphasize that for the evaluation of the AICc in this panel we used all of the model parameters, not just the number of unknown coefficients, $a_i$. 

The simple model
 with constant mass-to-light ratio and no separate dark matter component is favoured ($\text{AICc} = 86.50$). The second best is the one with a black hole in the centre ($\text{AICc} = 98.49$). 
For model inference with the AICc criterion, one is interested in  the difference of the AICc values with the best model (lowest value of AICc). In Fig \ref{DLI_III_FigFornax_2} we visualize these differences. The vertical axis is in log scale.   In Table  \ref{DLI_III_Fornax_AICc} we list all values of AICc as well as the differences $\Delta^i = \text{AICc}^i - \min (\text{AICc})$. Model comparison  is conclusive \citep{opac-b1100695}: the best model is a simple King profile with constant mass-to-light ratio. There is no need for the assumption of a  separate dark matter component with a significantly different mass profile. For the best  model, the knot distribution estimated from the EA  is uniform with only 3 knots, i.e. $\boldsymbol{\xi} = \{0.00, 0.50, 1.00\}$.    
Remarkably, despite the apparent flattening of Fornax that should increase the dispersion profile and perhaps suggest a different than the simple mass-follows-light DM component, our algorithm rejects such a scenario.  
The rejection of a cuspy profile is in accordance with the results of \cite{2013MNRAS.429L..89A} and \cite{2011ApJ...742...20W}.
 We estimate a total mass,   $M_{\text{tot}} = 1.613 ^{+0.050}_{-0.075} \times 10^8 \text{M}_{\odot}$ and a mass-to-light ratio, 
 $\Upsilon_V = 8.93 ^{+0.32}_{-0.47}\; \text{M}_{\odot}/\text{L}_{\odot}$. The best fitted King profile has parameters $ w_0 = 2.60^{+0.14}_{-0.28}$, $\rho_{0 \star } = 0.120^{+0.009}_{-0.004}\; \text{M}_{\odot} \text{pc}^{-3}$, $r_{\mathrm{c}} = 685^{+29}_{-38} \; \text{pc}$.   We summarize our results  in Table \ref{Fornax_best_params_via_mcmc}. 
 In Fig \ref{DLI_III_FigFornax_5} we plot (left panel) the marginalized distributions of mass-to-light ratio, and total mass. On the right panel we plot the normalized histogram of the marginalized total mass distribution. 
 In Table \ref{F_EA_sol}  we list the best estimates for the knots and parameters from the Evolutionary Algorithm.  The knots are divided by the tidal radius for each system and scaled in the interval $[0,1]$.  It needs to be emphasized that the tidal radius of each of the models depends on the adopted DM profile. This results from the solution of the Poisson equation for the King brightness profile. In all cases the tidal radius of the system, $r_{\mathrm{t}}$, is found to be much larger than the outermost radial bin of data.

\begin{figure*}
\centering
%height=0.3\textwidth, width=0.5\textwidth
%\showthe\columnwidth 
\includegraphics[width=\textwidth]{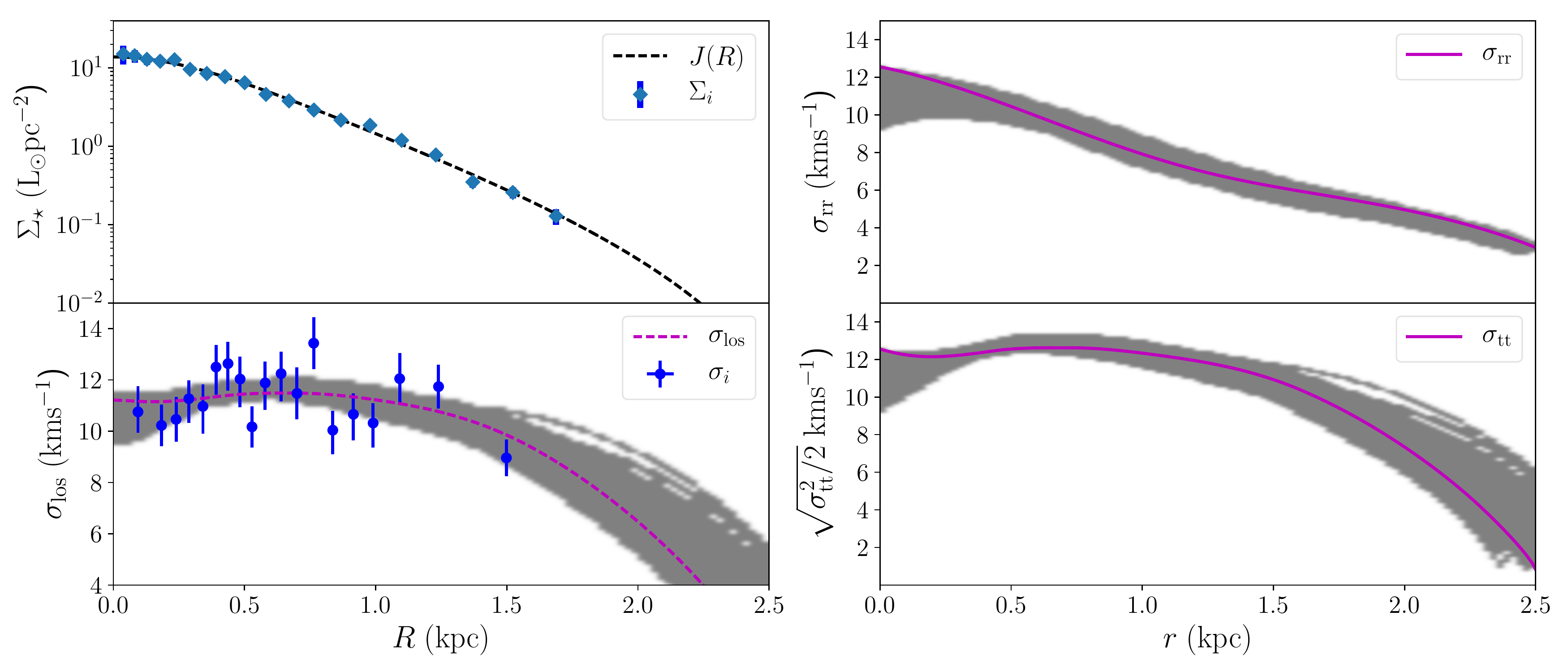}
\caption{Surface brightness and kinematic fits for the Fornax dSph.
Top left panel: highest likelihood surface brightness fit and data values. Bottom left panel: line-of-sight velocity dispersion values. Blue points correspond to observed values. The dotted purple line is the highest likelihood fit. Top right panel: radial velocity dispersion fit. Bottom right panel: tangential velocity dispersion, $\sigma_{\text{tt}}^2 = \sigma_{\theta\theta}^2 + \sigma_{\phi\phi}^2$. In all panels, the grey region corresponds to the $1\sigma$ uncertainty interval in all model parameters.}
\label{DLI_III_FigFornax_3}
\end{figure*}

 In Fig. \ref{DLI_III_FigFornax_3}, we plot in the top left panel the brightness fit. In the bottom left panel the line-of-sight velocity dispersion data values, $\sigma_{\text{los}}$, we estimated as well as the maximum likelihood fit for the best model (dashed line).  In the top right panel we plot the radial velocity dispersion, $\sigma_{\text{rr}}$, as this is recovered from our algorithm. In the bottom right panel the tangential, $\sqrt{\sigma_{\text{tt}}^2/2}$ component is shown.  In all figures  
the grey region corresponds to $1\sigma$ uncertainty in all parameters.

\begin{table}
\caption{Fornax AICc comparison of models.}
\label{DLI_III_Fornax_AICc}
\begin{center}
\begin{tabular}{ | l | l | l |  }
\hline
Model & $\text{AICc}$ & $\Delta$ \\ \hline
King, $\Upsilon_V$ &  $86.50$ & $0.0$ \\ 
King, $\Upsilon_V$, BH & $98.49$  & $11.99$ \\ 
King, $\Upsilon_V$, NFW & $112.66$  & $26.15$ \\ 
King, $\Upsilon_V$, Burkert & $114.36$ & $27.85$ \\ 
King, $\Upsilon_V$, Einasto & $135.38$ & $48.88$ \\ \hline
\end{tabular}
\end{center}
\end{table}

\begin{table}
\caption{MCMC estimated parameters for the best model for the case of Fornax.  The knots are scaled in the interval $[0,1]$. }
\begin{center}
\begin{tabular}{|l|l|l|}
\hline 
Best model    & Parameter &value \\ \hline
 &  $\boldsymbol{\xi}$  &$\{ 0 , 0.5, 1\}$\\[5pt]
& $ w_0$ &$ 2.60^{+0.14}_{-0.28}$
 \\[5pt]
 & $\rho_{0 \star }$ & $0.120^{+0.009}_{-0.004}\; \text{M}_{\odot} \text{pc}^{-3}$ \\[5pt]
const $\Upsilon_V$  &  $r_{\mathrm{c}}$  & $685^{+29}_{-38} \; \text{pc}$  \\[5pt]
& $\Upsilon_V$ & $  8.93 ^{+0.32}_{-0.47}\; \text{M}_{\odot}/\text{L}_{\odot}$ \\[5pt]
& $M_{\text{tot}}$ & $  1.613 ^{+0.050}_{-0.075} \times 10^8 \text{M}_{\odot} $\\[5pt]
& $\langle \beta \rangle$ & $-0.95^{+0.78}_{-0.72} $
\\ \hline 
 \end{tabular}
 \end{center}
\label{Fornax_best_params_via_mcmc}
\end{table}

\begin{table}
\caption{EA estimated parameters for the various models we used for the case of Fornax. The mass density constants $\rho_{0\bullet, \star}$ are evaluated in M$_{\odot}\text{pc}^{-3}$. The scaling radii $r_\text{c}, r_\text{s}, r_{\text{e}}$ in pc. The knots are scaled in the interval $[0,1]$.}
\begin{center}
\begin{tabular}{|l|l|l|l|}
\hline 
Model    & Parameter &value \\ \hline
const $\Upsilon_V$ &  $\boldsymbol{\xi}$  &$\{ 0 , 0.500 , 1\}$\\
& $\theta_{\star}=\{w_0, \rho_{0 \star} , r_{\mathrm{c}} \}$  &$\{ 2.537,  0.121, 691.613 \} $ \\
& $\Upsilon_V$ & $ 8.968 \; \text{M}_{\odot}/\text{L}_{\odot}$\\ \hline 
BH &  $\boldsymbol{\xi}$  &$\{ 0 , 0.496 , 1\}$\\
& $\theta_{\star}=\{w_0, \rho_{0 \star} , r_{\mathrm{c}} \}$  &$\{ 3.535  ,0.388, 380.678 \} $ \\
& $\Upsilon_V$ & $ 8.968  \; \text{M}_{\odot}/\text{L}_{\odot}$\\ 
& $\theta_{\bullet}= \{\text{M}_{\bullet} \}$ & $7.0565 \times 10^4 \text{M}_{\odot}$ \\\hline 
NFW &  $\boldsymbol{\xi}$  &$\{0 , 0.125 , 1\}$\\
& $\theta_{\star}=\{w_0, \rho_{0 \star} , r_{\mathrm{c}} \}$  &$\{ 1.809,  0.173,  761.048\} $ \\
& $\Upsilon_V$ & $ 6.816 \; \text{M}_{\odot}/\text{L}_{\odot}$\\
& $\theta_{\bullet}= \{\rho_{0\bullet},r_{\mathrm{s}} \}$  & $\{0.311,  170.067 \}$\\ \hline
Burkert  & $\boldsymbol{\xi}$  &$\{0 , 0.123 , 1\}$\\
& $\theta_{\star}=\{w_0, \rho_{0 \star} , r_{\mathrm{c}} \}$  &$\{ 2.551,  0.126,  687.322\} $ \\
& $\Upsilon_V$ & $ 8.594 \; \text{M}_{\odot}/\text{L}_{\odot}$\\
& $\theta_{\bullet} = \{\rho_{0\bullet},r_{\mathrm{s}} \}$  & $\{ 0.042,  196.315 \}$\\ \hline  
Einasto &  $\boldsymbol{\xi}$  &$\{ 0 , 0.131 , 1\}$\\
& $\theta_{\star}=\{w_0, \rho_{0 \star} , r_{\mathrm{c}} \}$  &$\{ 2.263,  0.173  ,1077.640 \} $ \\
& $\Upsilon_V$ & $ 8.830 \; \text{M}_{\odot}/\text{L}_{\odot}$\\
& $\theta_{\bullet}= \{\rho_{e},r_{\mathrm{e}}, n \}$  & $\{0.007  ,1324.556  ,  2.278 \}$\\ \hline 
 \end{tabular}
 \end{center}
\label{F_EA_sol}
\end{table}

\begin{figure}
\centering
%height=0.3\textwidth, width=0.5\textwidth
%\showthe\columnwidth 
%\includegraphics[width=\columnwidth]{images/Fornax_anisotropy_sigmaLOS.pdf}
\includegraphics[width=\columnwidth]{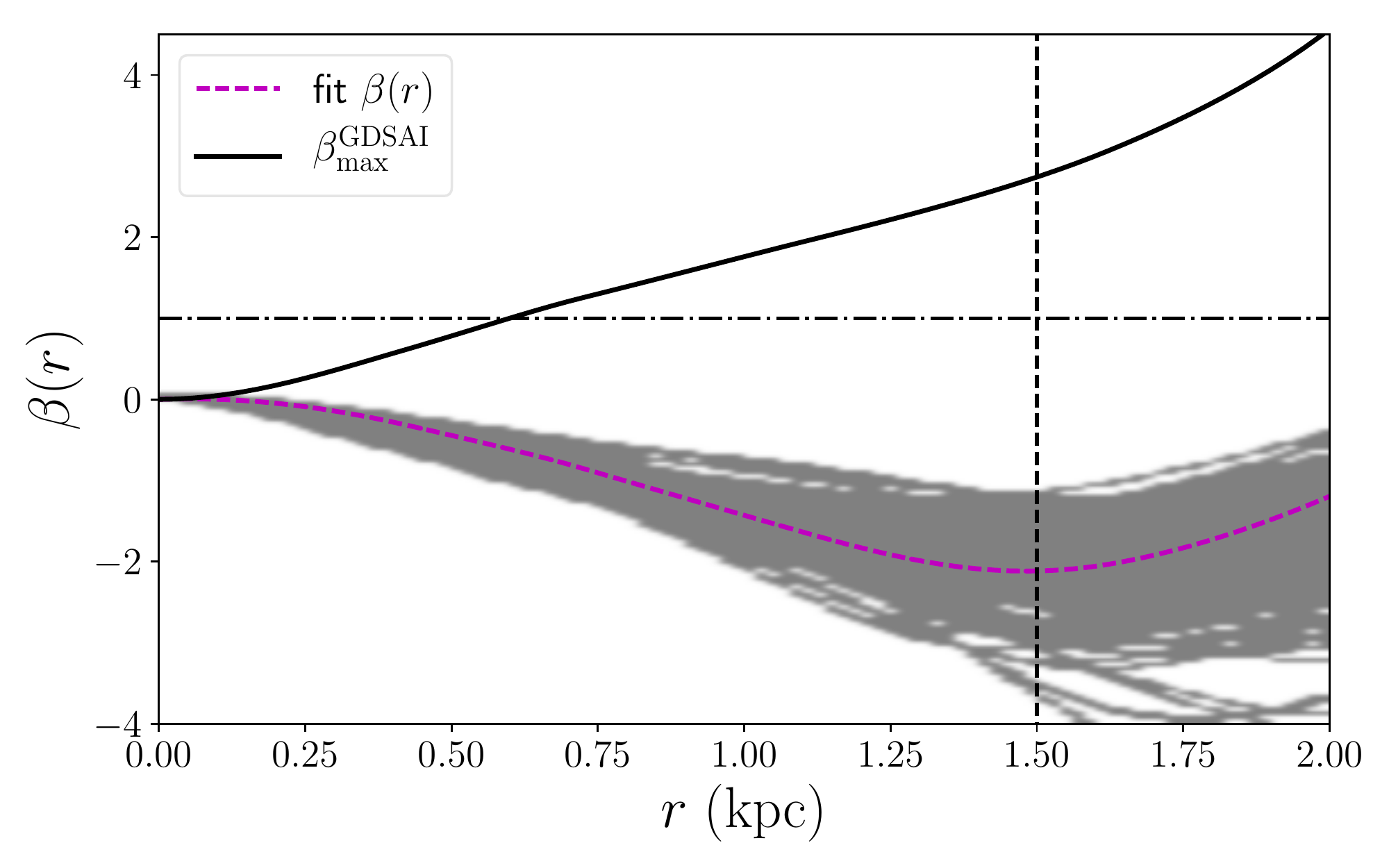}
\caption{Velocity anisotropy profile of the Fornax dSph. Grey region corresponds to $1\sigma$ uncertainty interval in all model parameters. The vertical dashed line marks the end point of our data, i.e. the limit of trusted predictions.  Beyond this point the profile is extrapolated and should not be trusted. The horizontal dashed-dot line corresponds to the physically limiting value $\beta = 1$. We also overplot the maximum value of the anisotropy, $\beta^{\mathrm{GDSAI}}_{\max}$ (solid black line),  as predicted according to the global density-slope anisotropy inequality.} 
\label{DLI_III_FigFornax_4}
\end{figure}

In Fig \ref{DLI_III_FigFornax_4} we plot the anisotropy profile. This starts from zero, in accordance with theoretical predictions: clearly, at $r=0$ tangential and radial motions are equivalent, thus\footnote{This can also be verified from the Jeans equation, Eq. \eqref{DLI_III_Jeans_std} in the limit $r\to 0$ for cored stellar profiles, $\rho_{\star}$ and a non-singular potential in the origin $\lim_{r\to 0}d\Phi/dr= 0$.} $\beta(0) = 0$. The anisotropy becomes progressively negative, with the uncertainty in the fit increasing. The vertical dashed line marks approximately the point where the range of our data ends, so anything beyond that is extrapolation and should not be trusted. In general, as stated previously,  our algorithm does not constrain well the velocity anisotropy, $\beta(r)$. However this has no effect in the modelling process we follow, since we do not use any $\beta(r)$ assumptions for our mass estimates. We also overplot the theoretical maximum possible value of the anisotropy, as this is predicted from the Global Density-Slope Anisotropy Inequality  \citep[][hereafter GDSAI\footnote{The GDSAI states that $\beta \leq  \gamma(r)/2$, where $\gamma(r) = -   \mathrm{d} \log \rho_{\star}(r) / \mathrm{d} \log (r)$.}, see also \citealt{2006ApJ...642..752A}]{2010MNRAS.408.1070C}. It should be stressed that the GDSAI is valid for systems where the stellar mass density is a separable function of radius and total potential, $\rho(r,\Phi)=A(r) B(\Phi)$. This is  the case for the King stellar profile in our  formulation. The anisotropy we calculate satisfies GDSAI in all distances, although after $\approx 0.6 \; \mathrm{kpc}$ it becomes meaningless since it goes above physically acceptable values ($\beta = 1$, dashed-dot horizontal line).

In order to perform a comparison with anisotropy estimates  of Fornax from other authors, we evaluate the average anisotropy, $\langle \beta \rangle$, from the origin up to $\sim 1.5$ kpc. We find $\langle \beta \rangle = -0.95^{+0.78}_{-0.72} $ within 1$\sigma$ uncertainty. Clearly the  average anisotropy according to the \cite{1980MNRAS.190..873B} definition is poorly constrained, however it lies within the range quoted by other authors \citep{2007ApJ...667L..53W,2009MNRAS.394L.102L,2013A&A...558A..35B}.

\begin{figure}
\centering
%height=0.3\textwidth, width=0.5\textwidth
%\showthe\columnwidth 
%\includegraphics[width=\columnwidth]{images/Fornax_anisotropy_sigmaLOS.pdf}
\includegraphics[width=\columnwidth]{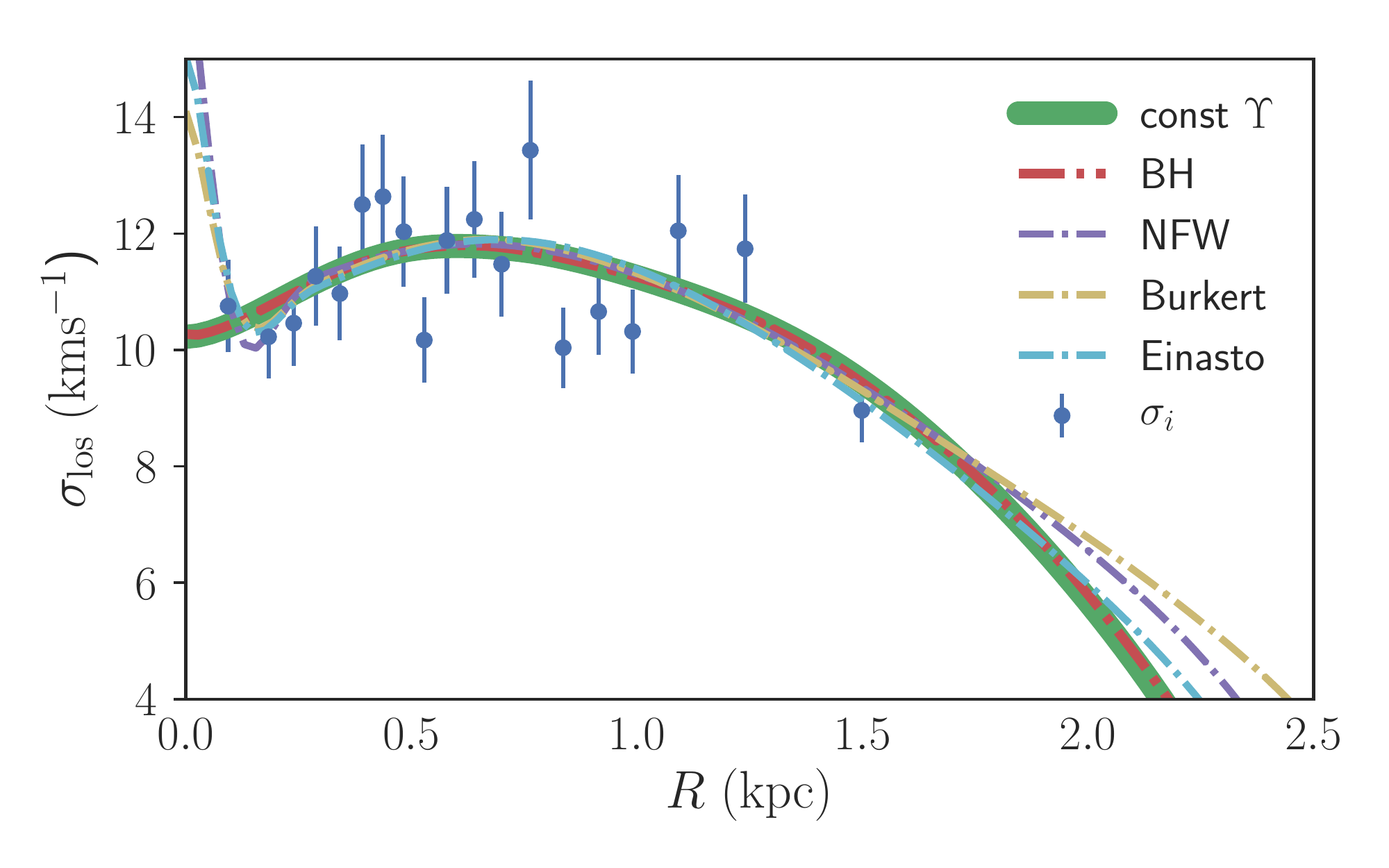}
\caption{Best EA $\sigma_{\mathrm{los}}$ profiles for all competing models for the Fornax dSph.}
\label{DLI_III_FigFornax_EA_all}
\end{figure}

Finally in order to understand why the EA chose to reject the models with a separate DM component, we plot in  Fig. \ref{DLI_III_FigFornax_EA_all} the $\sigma_{\mathrm{los}}$ profiles for the best EA selected competing  models. All of the models with a DM component have a sharp turn close to the origin. This is the reason the EA rejects them: they posses unnecessary complexity. In contrast the const$\Upsilon_V$ as well as the BH model have much smoother profiles close to the origin.

\subsubsection{Comparison with previous work}

\cite{2007ApJ...667L..53W} perform a simple NFW profile fit, with the assumption of constant anisotropy. They estimate a total mass of $M_{\mathrm{vir}} \sim 10^9\, \text{M}_{\odot}$, while the total mass within the maximum data point ($r_{\max}$) is $M_{r_{\text{max}}} = 1.8\times 10^8\, \text{M}_{\odot}$. Using $r_{\text{max}} \approx 1.8 \, \mathrm{(kpc)}$ we estimate $M_{r_{\text{max}} } \approx 1.549 \times  10^8 \, \text{M}_{\odot}$. While we are close to their $M_{r_{\text{max}}}$, we differ by orders of magnitude to their total (virial) mass estimate.  The authors fit a simple King (isotropic) profile with a constant mass-to-light ratio, and they exclude  it since it clearly fails to recover flat dispersion profiles. Quite remarkably, when we use a simple King profile, but with arbitrary anisotropy, we are able to recover the relatively flat line-of-sight velocity dispersion profile of Fornax. So a simple mass-follows-light model, with a non parametric anisotropic profile, is a viable candidate (Fig.  \ref{DLI_III_FigFornax_3}). This result is also in accordance with the suggestions of  \cite{2015MNRAS.454.2401B} that ``baryons and DM may contribute similarly to the mass budget in Fornax's central regions''.

\cite{2007MNRAS.378..353K} modelled Fornax by assuming a simple mass-follows-light system, a constant anisotropy and by  using the method of interlopers removal \citep{1996MNRAS.279..349D}. 
The authors estimated an anisotropy 
$\beta = -0.17^{+0.37}_{-0.63}$, a mass-to-light ratio $M/L = 11.3^{+2.1}_{-1.8}\,\text{M}_{\odot}/\text{L}_{\odot}$ and mass $M=2.1 \times 10^8 \, \text{M}_{\odot}$. 
 They also performed  a simple $2.58 \sigma$ cut in the line-of-sight velocities, and again modelled the dwarf, finding a more tangential anisotropy , $\beta = -1.82^{+1.02}_{-2.66}$ and a smaller mass-to-light ratio $M/L = 10.6^{+1.8}_{-1.7}\, \text{M}_{\odot}/\text{L}_{\odot}$. 
The anisotropy and mass-to-light ratios are, within error bars, in close proximity to  our results and their mass estimate is of the same order of magnitude with our estimates.

\cite{2009MNRAS.394L.102L} modelled Fornax by assuming a simple mass-follows-light model,  a constant anisotropy and using information from the kurtosis of the line-of-sight velocities. She estimated a total mass $M_{\text{tot}} = (1.57 \pm 0.07 ) \times 10^8 \text{M}_{\odot}$ which is in close proximity with our estimates. Similar results hold also for her constant mass-to-light ratio, $\Upsilon_V$, which was estimated to be $\Upsilon_V = (8.8 \pm 3.8)\,  \text{M}_{\odot}/\text{L}_{\odot}$. Our predictions agree within the error bars, however our mass-to-light ratio is  more tightly constrained. Furthermore she also predicted a  tangential anisotropy but in her case it is constant $\beta = - 0.3 \pm 0.15 < 0$. In her work the author does not perform model inference between competing mass models.  

 \cite{2013MNRAS.429L..89A} used a multiple stellar population model. They estimated a scale radius for a cored (Burkert) DM profile of the order $r_{\mathrm{s}} = 1.0^{+0.8}_{-0.4} \, \mathrm{kpc}$. Our best fitting model (const$\Upsilon_V$, with no separate DM component) has a scale radius of $r_c = 0.685^{+0.029}_{-0.038} \, \mathrm{kpc}$, that clearly overlaps with the lower uncertainty limit of \cite{2013MNRAS.429L..89A}.

\cite{2012ApJ...746...89J} model Fornax using the  \cite{1979ApJ...232..236S} method.  They test three DM models, a NFW profile, and a logarithmic one with and without a black hole. They do not include in their models a simple mass-follows-light mass profile. 
They estimate $M_{300} = 3.5 ^{+0.77}_{-0.11}\times 10^7 \, \text{M}_{\odot}$ while we find $M_{300} \sim 1.13 \times 10 ^7 \, \text{M}_{\odot}$. They  base their model selection in $\chi^2$ comparison (which is a weak  model inference method)  and conclude that a cored model without a black hole is a better description of the dSph.

\cite{2013A&A...558A..35B} also use the \cite{1979ApJ...232..236S} method for modelling Fornax. They test various DM profiles, and calculate the Bayesian evidence for the most probable model. Their model selection does not give a clear distinction between cuspy or core profiles. They estimate a mass within $1$kpc radius of $M_{1\mathrm{kpc}} \sim 10^8 \, \text{M}_{\odot}$, which is in close proximity with our estimate at the same distance, $M_{\star + \bullet}(r=1\,\mathrm{kpc}) \approx 1.09 \times 10^8 \,\text{M}_{\odot}$. They estimate an average constant anisotropy, $\langle \beta \rangle \sim -0.2 \pm 0.2$. This range is consistent with the average  anisotropy we calculate.

\section{Conclusions}
\label{DLI_III_Conclusions}

In this section we  report our conclusions from this contribution. For the convenience of the readers, we separate them in conclusions for the \textsc{JEAnS} algorithm we developed, and for the case of the Fornax dSph.

\subsection{\jeans modelling}

Building on previous work \citep{2014MNRAS.443..598D,2014MNRAS.443..610D}, we improve our method of modelling spherical systems independent of  velocity anisotropy assumptions.     
For this, we develop an evolutionary  algorithm that evaluates the optimum knot distribution for the B-spline representation of $\sigma_{\text{rr}}^2$, i.e. the best kinematic profile independent of the anisotropy $\beta(r)$. In addition, the \jeans discriminates between competing mass models using the AICc model selection criterion.   
 This model inference criterion has the advantage that is fast to evaluate and it can be applied to small samples of data. It has the disadvantage that it does not include information from the whole range of Markov chains in the MCMC. As a result, it is not as robust as Bayesian inference methods, e.g. model inference using Bayesian evidence \citep{2009MNRAS.398.1601F}. In tests with synthetic data sets, our method performed well and managed to discriminate between competing DM models.  

We extend the notion of empirical Bayes \citep{1985}, by presenting a new algorithm for the evaluation of prior information from ideal theoretical models  for the optimum smoothing of $\sigma_{\mathrm{rr}}^2$ for the line-of-sight, $\sigma_{\text{los}}^2$,  fit. This new version results tighter constraints on the smoothing hyperprior parameters and is computationally much faster than the one presented in \cite{2014MNRAS.443..610D}.

Finally, with our Jeans modelling independent of assumptions on
velocity anisotropy, we are able to discriminate between different parametric mass models. Thus, we are able to model independent of the mass-anisotropy degeneracy.

\subsection{Fornax}

We apply our algorithm to  Fornax dSph galaxy. Our method uses AICc for model inference. Based on the available brightness and kinematic data sets, our algorithm predicts conclusively \citep[$\Delta(\text{AICc}) > 20$,][]{opac-b1100695} that there is no need for a  separate dark matter component in the dwarf galaxy. 
That is, from a variety of cored and cuspy DM profiles and modelling independent of the MAD our best candidate  is a simple mass-follows-light model. 
  This does not imply that there is no dark matter in the dSph, however  
it does suggest that in a well mixed system, like Fornax, there is no need for a separate DM component that does not follow the stellar profile. 
  The second best candidate, which is also strongly disfavoured ($\Delta(\text{AICc}) \approx 12$), is a simple mass-follows-light model with a black hole in the centre. 
We  emphasize that these results should be verified with the use of proper Bayesian inference and the use of different tracer profiles;  this is currently a work in progress. 
 
We estimate a mass of $M_{\text{tot}} = 1.613 ^{+0.050}_{-0.075}\times 10^8 \, \text{M}_{\odot}$ and a mass-to-light ratio, 
 $\Upsilon_V = 8.93 ^{+0.32}_{-0.47} \, \text{M}_{\odot}/\text{L}_{\odot}$. These  values are consistent with previous work 
 \citep{2007ApJ...667L..53W, 2009MNRAS.394L.102L,2012ApJ...746...89J, 2013A&A...558A..35B}, however our mass-to-light ratio is more tightly constrained. 
 Considering that we did not use the best photometric data available for the brightness profile and that we did not account for the  non-negligible ellipticity $\epsilon \approx 0.3$ of Fornax, we expect that the value of the mass-to-light ratio, and the total mass of the galaxy can be pushed  to lower values.

 We find an anisotropy profile that is tangentially biased.   
 This tangential bias increases towards the outer parts of the dwarf,  
 however our algorithm cannot tightly constrain the $\beta(r)$ anisotropy. This is because we fit directly for $\sigma_{\text{rr}}^2$ and when we estimate $\beta(r)$ with the use of the Jeans equation, error propagation is significant. 
 The tangential anisotropy seems to favour the scenario that Fornax has undergone a recent minor merger \citep{2007MNRAS.378..353K,2041-8205-756-1-L18}. \cite{2041-8205-756-1-L18} performed numerical simulations for Fornax and argued that an initially disky galaxy can be transformed through tidal stirring to a spheroid. Thus this tangential anisotropy is perhaps a signature of an initial disky structure. \cite{ 2004ASPC..327..173C} observed a shell structure that could favour this scenario,  however this  is most likely  
 an incorrect assumption, resulting from a mis-identified overdensity of background galaxies \citep{2015MNRAS.453..690B}.

\section{Future direction}
\label{DLI_III_future}

The method we presented here has great potential for further  applications. Some of our goals for our future work is to expand our modelling for triaxial systems.  Other work in progress is to verify our  method with outputs from numerical simulations (Diakogiannis et al. in prep.) and finally apply our algorithm to the classical dwarfs to perform mass modelling independent of the MAD.

\section*{Acknowledgments}
 We wish to thank our referee, Gary Mamon, for his many useful comments and suggestions that helped us improve the paper.  Special thanks to  Matthias Redlich for providing a cubic spline interpolation  library, and  Dan Taranu for careful reading of the manuscript.   
F. I. Diakogiannis acknowledges support from ARC Discovery Project (DP140100198) and the Research Collaboration Awards (PG12105204). 
G. F. Lewis acknowledges support from ARC Discovery Project (DP110100678) and Future Fellowship (FT100100268).

%
%\nocite{*}
%\bibliographystyle{plain}
%\bibliography{references}
%\nocite{*} 
%\begin{thebibliography}{99}
\bibliographystyle{mn2e} %-> the bibliography style for MNRAS
\bibliography{DLI_III_V2} %-> YourListOfArticle is the compiled version of %YourListOfArticle.bib 

\begin{thebibliography}{}

\bibitem[\protect\citeauthoryear{{Amorisco}, {Agnello} \& {Evans}}{{Amorisco}
  et~al.}{2013}]{2013MNRAS.429L..89A}
{Amorisco} N.~C.,  {Agnello} A.,    {Evans} N.~W.,  2013, \mnras, 429, L89

\bibitem[\protect\citeauthoryear{{An} \& {Evans}}{{An} \&
  {Evans}}{2006}]{2006ApJ...642..752A}
{An} J.~H.,  {Evans} N.~W.,  2006, \apj, 642, 752

\bibitem[\protect\citeauthoryear{{Bate}, {McMonigal}, {Lewis}, {Irwin},
  {Gonzalez-Solares}, {Shanks} \& {Metcalfe}}{{Bate}
  et~al.}{2015}]{2015MNRAS.453..690B}
{Bate} N.~F.,  {McMonigal} B.,  {Lewis} G.~F.,  {Irwin} M.~J.,
  {Gonzalez-Solares} E.,  {Shanks} T.,    {Metcalfe} N.,  2015, \mnras, 453,
  690

\bibitem[\protect\citeauthoryear{{Battaglia}, {Sollima} \&
  {Nipoti}}{{Battaglia} et~al.}{2015}]{2015MNRAS.454.2401B}
{Battaglia} G.,  {Sollima} A.,    {Nipoti} C.,  2015, \mnras, 454, 2401

\bibitem[\protect\citeauthoryear{{Bicknell}, {Bruce}, {Carter} \&
  {Killeen}}{{Bicknell} et~al.}{1989}]{1989ApJ...336..639B}
{Bicknell} G.~V.,  {Bruce} T.~E.~G.,  {Carter} D.,    {Killeen} N.~E.~B.,
  1989, \apj, 336, 639

\bibitem[\protect\citeauthoryear{{Binney}}{{Binney}}{1980}]{1980MNRAS.190..873B}
{Binney} J.,  1980, \mnras, 190, 873

\bibitem[\protect\citeauthoryear{{Binney} \& {Mamon}}{{Binney} \&
  {Mamon}}{1982}]{1982MNRAS.200..361B}
{Binney} J.,  {Mamon} G.~A.,  1982, \mnras, 200, 361

\bibitem[\protect\citeauthoryear{{Binney} \& {Tremaine}}{{Binney} \&
  {Tremaine}}{2008}]{2008gady.book.....B}
{Binney} J.,  {Tremaine} S.,  2008, {Galactic Dynamics: Second Edition}.
Princeton University Press

\bibitem[\protect\citeauthoryear{Braak}{Braak}{2006}]{Braak:2006:MCM:1145406.1145416}
Braak C.~J.,  2006, Statistics and Computing, 16, 239

\bibitem[\protect\citeauthoryear{{Breddels} \& {Helmi}}{{Breddels} \&
  {Helmi}}{2013}]{2013A&A...558A..35B}
{Breddels} M.~A.,  {Helmi} A.,  2013, \aap, 558, A35

\bibitem[\protect\citeauthoryear{{Burkert}}{{Burkert}}{1995}]{1995ApJ...447L..25B}
{Burkert} A.,  1995, \apjl, 447, L25

\bibitem[\protect\citeauthoryear{Burnham \& Anderson}{Burnham \&
  Anderson}{2002}]{opac-b1100695}
Burnham K.~P.,  Anderson D. R.~b.,  2002, Model selection and multimodel
  inference : a practical information-theoretic approach.
Springer, New York

\bibitem[\protect\citeauthoryear{Casella}{Casella}{1985}]{1985}
Casella G.,  1985, The American Statistician, 39, pp. 83

\bibitem[\protect\citeauthoryear{{Ciotti} \& {Morganti}}{{Ciotti} \&
  {Morganti}}{2010}]{2010MNRAS.408.1070C}
{Ciotti} L.,  {Morganti} L.,  2010, \mnras, 408, 1070

\bibitem[\protect\citeauthoryear{Coello, Lamont \& Veldhuizen}{Coello
  et~al.}{2006}]{Coello:2006:EAS:1215640}
Coello C. A.~C.,  Lamont G.~B.,    Veldhuizen D. A.~V.,  2006, Evolutionary
  Algorithms for Solving Multi-Objective Problems (Genetic and Evolutionary
  Computation).
Springer-Verlag New York, Inc., Secaucus, NJ, USA

\bibitem[\protect\citeauthoryear{{Coleman}, {da Costa},
  {Mart{\'{\i}}nez-Delgado} \& {Bland-Hawthorn}}{{Coleman}
  et~al.}{2004}]{2004ASPC..327..173C}
{Coleman} M.,  {da Costa} G.~S.,  {Mart{\'{\i}}nez-Delgado} D.,
  {Bland-Hawthorn} J.,  2004, in {Prada} F.,  {Martinez Delgado} D.,
  {Mahoney} T.~J.,  eds,  Astronomical Society of the Pacific Conference Series
  Vol. 327, Satellites and Tidal Streams. p.~173

\bibitem[\protect\citeauthoryear{{Coleman}, {Da Costa}, {Bland-Hawthorn} \&
  {Freeman}}{{Coleman} et~al.}{2005}]{2005AJ....129.1443C}
{Coleman} M.~G.,  {Da Costa} G.~S.,  {Bland-Hawthorn} J.,    {Freeman} K.~C.,
  2005, \aj, 129, 1443

\bibitem[\protect\citeauthoryear{{Courteau} et~al.,}{{Courteau}
  et~al.}{2014}]{2014RvMP...86...47C}
{Courteau} S.  et~al., 2014, Reviews of Modern Physics, 86, 47

\bibitem[\protect\citeauthoryear{{de Blok}, {McGaugh} \& {Rubin}}{{de Blok}
  et~al.}{2001}]{2001AJ....122.2396D}
{de Blok} W.~J.~G.,  {McGaugh} S.~S.,    {Rubin} V.~C.,  2001, \aj, 122, 2396

\bibitem[\protect\citeauthoryear{Deb, Anand \& Joshi}{Deb
  et~al.}{2002}]{Deb:2002:CEE:638598.638601}
Deb K.,  Anand A.,    Joshi D.,  2002, Evol. Comput., 10, 371

\bibitem[\protect\citeauthoryear{{Dejonghe} \& {Merritt}}{{Dejonghe} \&
  {Merritt}}{1992}]{1992ApJ...391..531D}
{Dejonghe} H.,  {Merritt} D.,  1992, \apj, 391, 531

\bibitem[\protect\citeauthoryear{{den Hartog} \& {Katgert}}{{den Hartog} \&
  {Katgert}}{1996}]{1996MNRAS.279..349D}
{den Hartog} R.,  {Katgert} P.,  1996, \mnras, 279, 349

\bibitem[\protect\citeauthoryear{{Diakogiannis}, {Lewis} \&
  {Ibata}}{{Diakogiannis} et~al.}{2014a}]{2014MNRAS.443..598D}
{Diakogiannis} F.~I.,  {Lewis} G.~F.,    {Ibata} R.~A.,  2014a, \mnras, 443,
  598

\bibitem[\protect\citeauthoryear{{Diakogiannis}, {Lewis} \&
  {Ibata}}{{Diakogiannis} et~al.}{2014b}]{2014MNRAS.443..610D}
{Diakogiannis} F.~I.,  {Lewis} G.~F.,    {Ibata} R.~A.,  2014b, \mnras, 443,
  610

\bibitem[\protect\citeauthoryear{{Einasto} \& {Haud}}{{Einasto} \&
  {Haud}}{1989}]{1989A&A...223...89E}
{Einasto} J.,  {Haud} U.,  1989, \aap, 223, 89

\bibitem[\protect\citeauthoryear{{Evans}, {An} \& {Walker}}{{Evans}
  et~al.}{2009}]{2009MNRAS.393L..50E}
{Evans} N.~W.,  {An} J.,    {Walker} M.~G.,  2009, \mnras, 393, L50

\bibitem[\protect\citeauthoryear{{Feroz}, {Hobson} \& {Bridges}}{{Feroz}
  et~al.}{2009}]{2009MNRAS.398.1601F}
{Feroz} F.,  {Hobson} M.~P.,    {Bridges} M.,  2009, \mnras, 398, 1601

\bibitem[\protect\citeauthoryear{Ferreau, Kirches, Potschka, Bock \&
  Diehl}{Ferreau et~al.}{2014}]{Ferreau2014}
Ferreau H.,  Kirches C.,  Potschka A.,  Bock H.,    Diehl M.,  2014,
  Mathematical Programming Computation, 6, 327

\bibitem[\protect\citeauthoryear{Foreman-Mackey et~al.,}{Foreman-Mackey
  et~al.}{2016}]{dan_foreman_mackey_2016_45906}
Foreman-Mackey D.  et~al.,, 2016, corner.py: corner.py v1.0.2

\bibitem[\protect\citeauthoryear{Ghosh, Dehuri \& Ghosh}{Ghosh
  et~al.}{2008}]{ghosh2008multi}
Ghosh A.,  Dehuri S.,    Ghosh S.,  2008, Multi-Objective Evolutionary
  Algorithms for Knowledge Discovery from Databases.
Studies in Computational Intelligence, Springer Berlin Heidelberg

\bibitem[\protect\citeauthoryear{{Gilmore}, {Wilkinson}, {Wyse}, {Kleyna},
  {Koch}, {Evans} \& {Grebel}}{{Gilmore} et~al.}{2007}]{2007ApJ...663..948G}
{Gilmore} G.,  {Wilkinson} M.~I.,  {Wyse} R.~F.~G.,  {Kleyna} J.~T.,  {Koch}
  A.,  {Evans} N.~W.,    {Grebel} E.~K.,  2007, \apj, 663, 948

\bibitem[\protect\citeauthoryear{Goldberg}{Goldberg}{1989}]{Goldberg:1989:GAS:534133}
Goldberg D.~E.,  1989, Genetic Algorithms in Search, Optimization and Machine
  Learning, 1st edn.
Addison-Wesley Longman Publishing Co., Inc., Boston, MA, USA

\bibitem[\protect\citeauthoryear{{Goodman} \& {Weare}}{{Goodman} \&
  {Weare}}{2010}]{GoodmanWeare}
{Goodman} G.~G.,  {Weare} J.~J.,  2010, CAMCoS, 5, 65

\bibitem[\protect\citeauthoryear{Hansen, Auger, Ros, Finck \&
  Po\v{s}\'{\i}k}{Hansen et~al.}{2010}]{Hansen:2010:CRA:1830761.1830790}
Hansen N.,  Auger A.,  Ros R.,  Finck S.,    Po\v{s}\'{\i}k P.,  2010, in
  Proceedings of the 12th Annual Conference Companion on Genetic and
  Evolutionary Computation. GECCO '10.
ACM, New York, NY, USA, pp 1689--1696

\bibitem[\protect\citeauthoryear{Hansen, Hansen, Ostermeier \&
  Ostermeier}{Hansen et~al.}{1996}]{Hansen96adaptingarbitrary}
Hansen N.,  Hansen N.,  Ostermeier A.,    Ostermeier A.,  1996. Morgan
  Kaufmann, pp 312--317

\bibitem[\protect\citeauthoryear{Hastie, Tibshirani \& Friedman}{Hastie
  et~al.}{2001}]{hastie01statisticallearning}
Hastie T.,  Tibshirani R.,    Friedman J.,  2001, The Elements of Statistical
  Learning.
Springer Series in Statistics, Springer New York Inc., New York, NY, USA

\bibitem[\protect\citeauthoryear{Hurvich \& Tsai}{Hurvich \&
  Tsai}{1991}]{hurvich1991bias}
Hurvich C.~M.,  Tsai C.-L.,  1991, Biometrika, 78, 499

\bibitem[\protect\citeauthoryear{{Ibata}, {Nipoti}, {Sollima}, {Bellazzini},
  {Chapman} \& {Dalessandro}}{{Ibata} et~al.}{2013}]{2013MNRAS.428.3648I}
{Ibata} R.,  {Nipoti} C.,  {Sollima} A.,  {Bellazzini} M.,  {Chapman} S.~C.,
  {Dalessandro} E.,  2013, \mnras, 428, 3648

\bibitem[\protect\citeauthoryear{{Irwin} \& {Hatzidimitriou}}{{Irwin} \&
  {Hatzidimitriou}}{1995}]{1995MNRAS.277.1354I}
{Irwin} M.,  {Hatzidimitriou} D.,  1995, \mnras, 277, 1354

\bibitem[\protect\citeauthoryear{{Jardel} \& {Gebhardt}}{{Jardel} \&
  {Gebhardt}}{2012}]{2012ApJ...746...89J}
{Jardel} J.~R.,  {Gebhardt} K.,  2012, \apj, 746, 89

\bibitem[\protect\citeauthoryear{{King}}{{King}}{1966}]{1966AJ.....71...64K}
{King} I.~R.,  1966, \aj, 71, 64

\bibitem[\protect\citeauthoryear{{Klimentowski}, {{\L}okas}, {Kazantzidis},
  {Prada}, {Mayer} \& {Mamon}}{{Klimentowski}
  et~al.}{2007}]{2007MNRAS.378..353K}
{Klimentowski} J.,  {{\L}okas} E.~L.,  {Kazantzidis} S.,  {Prada} F.,  {Mayer}
  L.,    {Mamon} G.~A.,  2007, \mnras, 378, 353

\bibitem[\protect\citeauthoryear{{{\L}okas}}{{{\L}okas}}{2002}]{2002MNRAS.333..697L}
{{\L}okas} E.~L.,  2002, \mnras, 333, 697

\bibitem[\protect\citeauthoryear{{{\L}okas}}{{{\L}okas}}{2009}]{2009MNRAS.394L.102L}
{{\L}okas} E.~L.,  2009, \mnras, 394, L102

\bibitem[\protect\citeauthoryear{{Mamon}, {Biviano} \& {Bou{\'e}}}{{Mamon}
  et~al.}{2013}]{2013MNRAS.429.3079M}
{Mamon} G.~A.,  {Biviano} A.,    {Bou{\'e}} G.,  2013, \mnras, 429, 3079

\bibitem[\protect\citeauthoryear{{Mamon} \& {Bou{\'e}}}{{Mamon} \&
  {Bou{\'e}}}{2010}]{2010MNRAS.401.2433M}
{Mamon} G.~A.,  {Bou{\'e}} G.,  2010, \mnras, 401, 2433

\bibitem[\protect\citeauthoryear{{Mamon} \& {{\L}okas}}{{Mamon} \&
  {{\L}okas}}{2005}]{2005MNRAS.362...95M}
{Mamon} G.~A.,  {{\L}okas} E.~L.,  2005, \mnras, 362, 95

\bibitem[\protect\citeauthoryear{{Mateo}}{{Mateo}}{1998}]{1998ARA&A..36..435M}
{Mateo} M.~L.,  1998, \araa, 36, 435

\bibitem[\protect\citeauthoryear{{Merritt}}{{Merritt}}{1985}]{1985AJ.....90.1027M}
{Merritt} D.,  1985, \aj, 90, 1027

\bibitem[\protect\citeauthoryear{{Merritt}}{{Merritt}}{1987}]{1987ApJ...313..121M}
{Merritt} D.,  1987, \apj, 313, 121

\bibitem[\protect\citeauthoryear{{Merritt}}{{Merritt}}{2013}]{2013degn.book.....M}
{Merritt} D.,  2013, {Dynamics and Evolution of Galactic Nuclei}

\bibitem[\protect\citeauthoryear{{Merritt}, {Graham}, {Moore}, {Diemand} \&
  {Terzi{\'c}}}{{Merritt} et~al.}{2006}]{2006AJ....132.2685M}
{Merritt} D.,  {Graham} A.~W.,  {Moore} B.,  {Diemand} J.,    {Terzi{\'c}} B.,
  2006, \aj, 132, 2685

\bibitem[\protect\citeauthoryear{Mäkinen, Periaux \& Toivanen}{Mäkinen
  et~al.}{1999}]{FLD:FLD829}
Mäkinen R.~A.,  Periaux J.,    Toivanen J.,  1999, International Journal for
  Numerical Methods in Fluids, 30, 149

\bibitem[\protect\citeauthoryear{{Navarro}, {Frenk} \& {White}}{{Navarro}
  et~al.}{1996}]{1996ApJ...462..563N}
{Navarro} J.~F.,  {Frenk} C.~S.,    {White} S.~D.~M.,  1996, \apj, 462, 563

\bibitem[\protect\citeauthoryear{{Navarro} et~al.,}{{Navarro}
  et~al.}{2004}]{2004MNRAS.349.1039N}
{Navarro} J.~F.  et~al., 2004, \mnras, 349, 1039

\bibitem[\protect\citeauthoryear{{Osipkov}}{{Osipkov}}{1979}]{1979SvAL....5...42O}
{Osipkov} L.~P.,  1979, Soviet Astronomy Letters, 5, 42

\bibitem[\protect\citeauthoryear{{Read} \& {Steger}}{{Read} \&
  {Steger}}{2017}]{2017arXiv170104833R}
{Read} J.~I.,  {Steger} P.,  2017, ArXiv e-prints

\bibitem[\protect\citeauthoryear{{Richardson} \& {Fairbairn}}{{Richardson} \&
  {Fairbairn}}{2013}]{2013MNRAS.432.3361R}
{Richardson} T.,  {Fairbairn} M.,  2013, \mnras, 432, 3361

\bibitem[\protect\citeauthoryear{Rozenberg, Bck \& Kok}{Rozenberg
  et~al.}{2011}]{Rozenberg:2011:HNC:2016676}
Rozenberg G.,  Bck T.,    Kok J.~N.,  2011, Handbook of Natural Computing, 1st
  edn.
Springer Publishing Company, Incorporated

\bibitem[\protect\citeauthoryear{{Schwarzschild}}{{Schwarzschild}}{1979}]{1979ApJ...232..236S}
{Schwarzschild} M.,  1979, \apj, 232, 236

\bibitem[\protect\citeauthoryear{Sugiura}{Sugiura}{1978}]{sugiura1978further}
Sugiura N.,  1978, Communications in Statistics-Theory and Methods, 7, 13

\bibitem[\protect\citeauthoryear{Talbi}{Talbi}{2009}]{Talbi:2009:MDI:1718024}
Talbi E.-G.,  2009, Metaheuristics: From Design to Implementation.
Wiley Publishing

\bibitem[\protect\citeauthoryear{{Tiret}, {Combes}, {Angus}, {Famaey} \&
  {Zhao}}{{Tiret} et~al.}{2007}]{2007A&A...476L...1T}
{Tiret} O.,  {Combes} F.,  {Angus} G.~W.,  {Famaey} B.,    {Zhao} H.~S.,  2007,
  \aap, 476, L1

\bibitem[\protect\citeauthoryear{{Walcher}, {Fried}, {Burkert} \&
  {Klessen}}{{Walcher} et~al.}{2003}]{2003A&A...406..847W}
{Walcher} C.~J.,  {Fried} J.~W.,  {Burkert} A.,    {Klessen} R.~S.,  2003,
  \aap, 406, 847

\bibitem[\protect\citeauthoryear{{Walker}, {Mateo} \& {Olszewski}}{{Walker}
  et~al.}{2009}]{2009AJ....137.3100W}
{Walker} M.~G.,  {Mateo} M.,    {Olszewski} E.~W.,  2009, \aj, 137, 3100

\bibitem[\protect\citeauthoryear{{Walker}, {Mateo}, {Olszewski}, {Gnedin},
  {Wang}, {Sen} \& {Woodroofe}}{{Walker} et~al.}{2007}]{2007ApJ...667L..53W}
{Walker} M.~G.,  {Mateo} M.,  {Olszewski} E.~W.,  {Gnedin} O.~Y.,  {Wang} X.,
  {Sen} B.,    {Woodroofe} M.,  2007, \apjl, 667, L53

\bibitem[\protect\citeauthoryear{{Walker}, {Mateo}, {Olszewski}, {Sen} \&
  {Woodroofe}}{{Walker} et~al.}{2009}]{2009AJ....137.3109W}
{Walker} M.~G.,  {Mateo} M.,  {Olszewski} E.~W.,  {Sen} B.,    {Woodroofe} M.,
  2009, \aj, 137, 3109

\bibitem[\protect\citeauthoryear{{Walker} \& {Pe{\~n}arrubia}}{{Walker} \&
  {Pe{\~n}arrubia}}{2011}]{2011ApJ...742...20W}
{Walker} M.~G.,  {Pe{\~n}arrubia} J.,  2011, \apj, 742, 20

\bibitem[\protect\citeauthoryear{{Weinberg}, {Bullock}, {Governato}, {Kuzio de
  Naray} \& {Peter}}{{Weinberg} et~al.}{2015}]{2015PNAS..11212249W}
{Weinberg} D.~H.,  {Bullock} J.~S.,  {Governato} F.,  {Kuzio de Naray} R.,
  {Peter} A.~H.~G.,  2015, Proceedings of the National Academy of Science, 112,
  12249

\bibitem[\protect\citeauthoryear{{Wolf}}{{Wolf}}{2010}]{2010HiA....15...79W}
{Wolf} J.,  2010, Highlights of Astronomy, 15, 79

\bibitem[\protect\citeauthoryear{{Wolf}}{{Wolf}}{2011}]{2011IAUS..271..110W}
{Wolf} J.,  2011, in {Brummell} N.~H.,  {Brun} A.~S.,  {Miesch} M.~S.,
  {Ponty} Y.,  eds,  IAU Symposium Vol. 271, IAU Symposium. pp 110--118

\bibitem[\protect\citeauthoryear{Yozin \& Bekki}{Yozin \&
  Bekki}{2012}]{2041-8205-756-1-L18}
Yozin C.,  Bekki K.,  2012, The Astrophysical Journal Letters, 756, L18

\bibitem[\protect\citeauthoryear{Zhou, Qu, Li, Zhao, Suganthan \& Zhang}{Zhou
  et~al.}{2011}]{journals/swevo/ZhouQLZSZ11}
Zhou A.,  Qu B.-Y.,  Li H.,  Zhao S.-Z.,  Suganthan P.~N.,    Zhang Q.,  2011,
  Swarm and Evolutionary Computation, 1, 32

\end{thebibliography}
%\end{thebibliography}

\appendix

\section{The importance of optimum knot distribution in B-spline fitting}
\label{DLI_III_simple_curve_Example}

\begin{figure}
\centering
%height=0.3\textwidth, width=0.5\textwidth
%\showthe\columnwidth 
\includegraphics[width=\columnwidth]{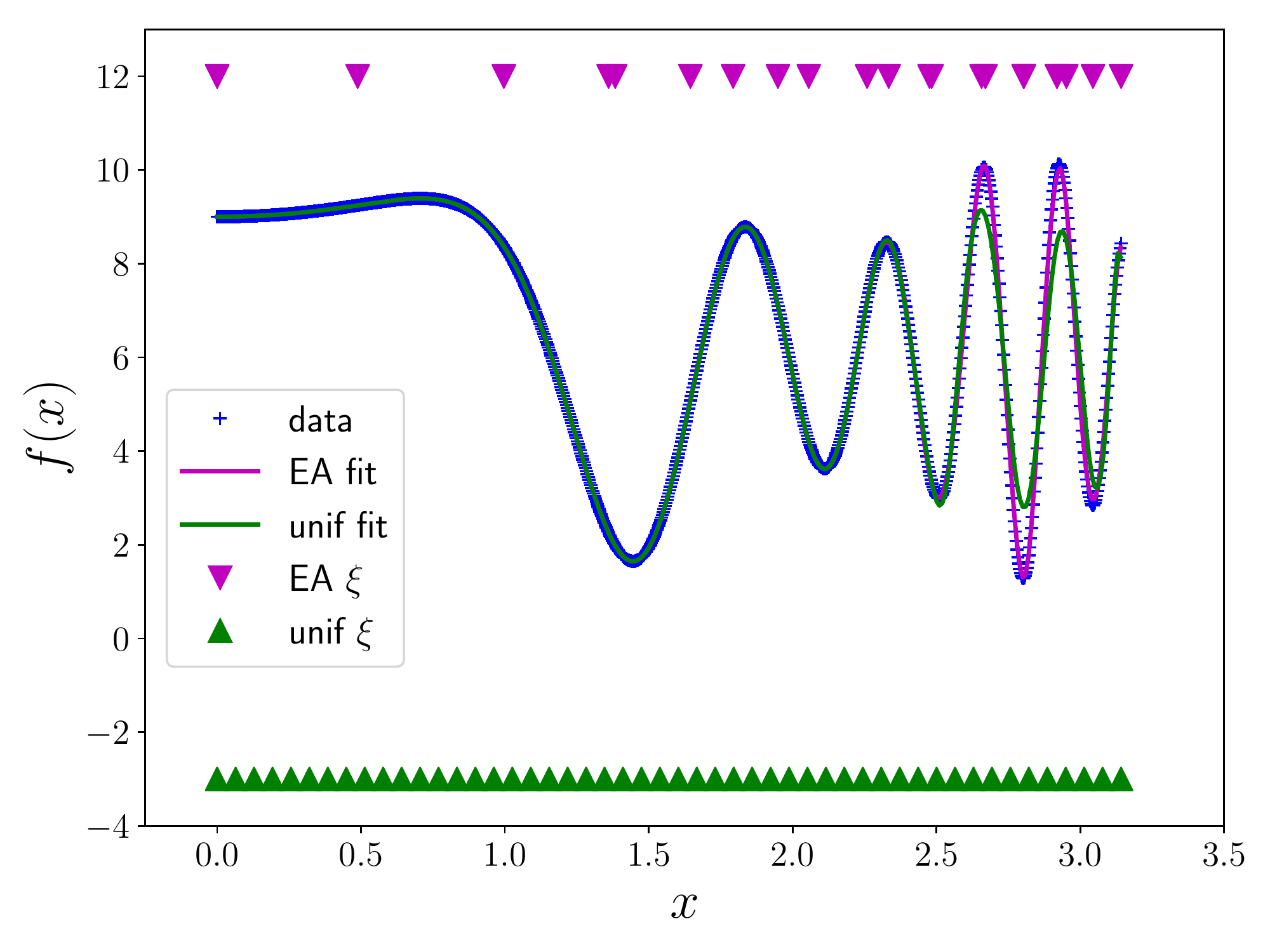}
\caption{Fit to synthetic data of the simple curve given by Eq. \eqref{DLI_III_toy_function_Appendix}. Blue crosses are the synthetic data. The purple line is the best EA recovered fit (24 knots). Green line is a fit with 50 uniform knots. Triangles pointing downwards correspond to the location of the knots, $\xi_i$ of the EA evaluation, while triangles pointing upwards, are the locations of the uniform knots. Clearly the EA places knots in regions where the data demonstrate higher curvature.}
\label{DLI_III_FigToy_appendix}
\end{figure}

In this section we  give a simple example of a curve fitting to  noiseless data in order to demonstrate the need for careful placement of knots.  We create 1000 data points, uniformly distributed from  a ``difficult'' toy function 
\begin{equation}
\label{DLI_III_toy_function_Appendix}
f(x) =6+\cos(x^3) (2+\exp\{\sin(x^2)\})
\end{equation}
We model this toy function by assuming the B-spline approximation of $f(x) \approx \sum_{i=1}^n a_i B_i(x)$. The total number, $n$, of unknown coefficients and the placement of the knots is evaluated with the use of the  EA. 
For simplicity we use  no smoothing penalty and no errors.  The EA  evaluates the best fit with 24 knots, $\xi$, appropriately placed with higher density in regions where the data demonstrate higher curvature. For comparison we also fit with a uniform distribution, with 50 knots. We visualize our results in Fig. \ref{DLI_III_FigToy_appendix}. Blue crosses correspond to the synthetic data. The purple line is the EA best fit, the green line the fit with the uniform distribution of knots. We overplot the positions of the EA selected knots (upper purple triangles, pointing downwards) as well as the  the  uniform knots (lower green triangles pointing upwards). Clearly the EA places more knots in regions where the data have higher curvature. In these regions (in the interval $x\in[2.5,3.25]$) the uniform fit has the greatest deviation from the true curve. We need $\sim 100$ knots to get a satisfactory fit when using uniform knots, however this has a severe impact on model selection methods, where models with smaller number of unknowns are favoured.  In the case where we would have included errors, the fit of the uniform knot distribution would be worst: the fitted curve would tend to follow the noise of the data due to the large number of knots and that would be more evident in regions of low curvature.  
Simply put, the optimum knot distribution requires locally adaptive smoothing.

\section{Comparison of modelling with uniform knots}
\label{DLI_III_Fornax_uniform_knots}
In order to get a quantitative measure of the EA performance, we re-ran  the EA for the case of Fornax for two models: the King const $\Upsilon_V$ and the model with Burkert DM. We used a set of uniform knots, $ \boldsymbol{\xi} = \{0, 0.25, 0.5, 0.75, 1 \}$ scaled to $[0,1]$. The King model again is chosen as the best model. The difference in AICc values becomes: 
 $\Delta \text{AICc} = \text{AICc}_\text{Burk} - \text{AICc}_\text{King} =  135.62 - 97.75 = 37.87$, while with the EA selected knots it is  $\Delta \text{AICc} = 27.85$.   Clearly the differences in AICc between the knots chosen by the EA and the uniform ones  are non negligible (difference $\approx$ 10).  We  note that difference $\geq 10$ is conclusive for model selection when we use the AICc. 
 This means that the EA helps to distinguish quantitatively between competing mass models. 
It also demonstrates that for comparison of  models  with fixed knots the AICc difference can be conclusive while using the EA adaptive modelling, it may not. 
 
 Some more insight can be given by the   values of  the training ``error'',  
$\chi^2=-\ln \mathcal{L}$    
  (Eq. \ref{DLI_III_logLEA_trainError}). The King model gives 
 $ \chi^2_{\mathrm{EA}} = 39.0$ with EA selected knots, but with uniform knots it gives $\chi^2_{\mathrm{unif}} = 38.0$, a lower value. This is expected since the uniform knots we used include also the EA selected knots, but the basis is more flexible. That is, uniform knots  tend to overfit the data.  This becomes clear when we compare the  AICc values  of the EA selected knots and the uniform ones: for the uniform knots, $\mathrm{AICc}^{\mathrm{unif}} = 97.75$, while for the EA selected ones, $\mathrm{AICc}^{\mathrm{EA}} = 86.50$, therefore the EA selected model is better than the uniform knots one ($\mathrm{AICc}^{\mathrm{EA}} <  \mathrm{AICc}^{\mathrm{unif}}$). 
 The Burkert model gives $\chi^2_{\mathrm{EA}} = 38.36$ with EA selected knots, but $\chi^2_{\mathrm{unif}} = 36.82$ with uniform knots. In this case, the uniform knots do not include the knot $\xi \sim 0.122$ that the EA finds, however, the basis is more flexible and the model has a tendency to overfit.
 Similar arguments hold for comparing the Burkert models (EA selected and uniform knots) with their AICc values.

 Another  quantitative measure of the EA performance can be seen by the fact that when we modelled Fornax, the EA gave solutions with the same number of knots - but different locations. These different models give significantly different AICc values, despite having the same number of unknown coefficients, $a_i$. This is because the EA finds the best trade off between the knots (numbers and location) and training error $\chi^2$ with the use of AICc.  If we fix the number and location of  the knots, as we did in \cite{2014MNRAS.443..598D,2014MNRAS.443..610D}, we lose the information of this trade off.

\section{EA terminology}
\label{DLI_III_EA_glossary}
This is a  glossary of terms that is used in the EA literature for the aid of readers not familiar with EAs. It is by no means complete but it is focused in terminology we use in this work.    
\begin{description}
\item[\it Individual] is a candidate solution, i.e. a set of variables. Sometimes the word chromosome is also used. 
In the simplest case this is just a vector of real numbers, e.g $(x_1,x_2,\ldots,x_n)$ this need not be the general case though.  
An individual can be any structure we can use as a representation inside a computer program that  can be used to assign a score of quality (fitness value) to the individual.  
\item[\it Population] is a finite  collection of individuals. For example, for the fitness function $f(x,y) = x^2+y^2$ a population consists of a set of pair of values $\{(x_i,y_i)\}$, $i=1,\ldots,N_{\text{pop}}$, where $N_{\text{pop}}$ is the total number of candidate solutions.    
\item[\it Generation.] In each iteration of the evolutionary algorithm, the software performs some operations in order to process the current population and produce a new one. The population at a specific iteration is called generation. 
\item[\it Fitness] of the individual is the value of a scalar\footnote{It the case of multiobjective optimization this can also be a vector function.} function (fitness function) that can describe the quality of the solution. 
Individuals with highest fitness value describe a better fit (solution) to the problem. 
\item[\it Parents] are individuals that are chosen for creating new solutions from them.
\item[\it Crossover] is the process of combining information from two or more individuals (parents) and produce one or more new individuals (offspring).
\item[\it Mutation.] The process of altering randomly an individual in order to produce a new solution.   
\item[\it Recombination.] The process of crossover and mutation. 
\item[\it Offspring] are individuals that are produced after the process of recombination.  
\item[\it Selection] is the process of selecting individuals from the current generation that will be used as parents to produce new solutions (offspring)
\item[\it Evolutionary model] is the process that evolves the algorithm in each generation. This can be much more complicated than just selection and recombination. For example, we can have multiple offspring from one or more parents, we can assign an age to each individual that ceases to exist after a certain point etc.     
\end{description}

\section{CPU processing times for Fornax}

In the version of the algorithm used for this paper we had approximately the following times: 
\begin{itemize}
\item EA for a single model run:  $\sim$48 h. 
\item  For the optimum smoothing hyperparameters: $\sim$24h  for a single model run. 
\item For the MCMC  $\sim$48 h for the best model run.  
\end{itemize}
The processing of Fornax took $\sim 2$ weeks. We need to emphasize though that the Phase 2 and 3 were applied only once, that is for the best selected model from Phase 1. 
 The processing times may vary depending on how ``difficult'' (how steep is the $\sigma_\text{los}^2$ profile) is a model to be fitted, as we found from experience with the \textsc{GAIA challenge} data set. These times are on a single laptop \textsc{(HP ZBook)}.

\bsp

\label{lastpage}

\end{document}